\documentclass[a4paper,11pt]{article}

\usepackage[ruled]{algorithm2e}
\usepackage{amsmath}
\usepackage{amssymb}
\usepackage{booktabs}
\usepackage{cite}
\usepackage{courier}
\usepackage{hyperref}
\usepackage{listings}
\usepackage[noBBpl]{mathpazo}
\usepackage{subfig}
\usepackage{tikz}
\usepackage{url}
\usepackage{xspace}

\newcommand{\plgRaw}{PLG2}
\newcommand{\plg}{\plgRaw{}\xspace}

\newcommand{\SEQ}{\ensuremath{;}}
\newcommand{\AND}{\mathbin{\oplus}}
\newcommand{\XOR}{\mathbin{\otimes}}
\newcommand{\LOOP}{\circlearrowright}
\newcommand{\DOIn}[2]{\ensuremath{#1\swarrow^{#2}}}
\newcommand{\DOOut}[2]{\ensuremath{#1\nearrow^{#2}}}

\newcommand{\seq}[1]{\ensuremath{\mathbb{N}^+_{#1}}}

\DeclareMathOperator{\out}{out}
\DeclareMathOperator{\incoming}{in}
\DeclareMathOperator{\random}{rnd}

\SetCommentSty{mycommfont}
\SetKwComment{Comment}{$\triangleright$\ }{}
\SetKwInOut{Input}{Input}
\SetKwInOut{Output}{Output}

\lstdefinestyle{customPython}{
	belowcaptionskip=1\baselineskip,
	breaklines=true,
	frame=single,
	language=Python,
	showstringspaces=false,
	basicstyle=\footnotesize\ttfamily,
	keywordstyle=\bfseries\color{blue!40!black},
	commentstyle=\color{gray!80!black},
	identifierstyle=\color{black},
}

\lstdefinestyle{customJava}{
	belowcaptionskip=1\baselineskip,
	breaklines=true,
	frame=single,
	language=Java,
	showstringspaces=false,
	basicstyle=\footnotesize\ttfamily,
	keywordstyle=\bfseries\color{blue!40!black},
	commentstyle=\color{gray!80!black},
	identifierstyle=\color{black},
	stringstyle=\color{red},
}

\title{\plg: Multiperspective Processes Randomization and Simulation for Online and Offline Settings}
\author{Andrea Burattin \thanks{A. Burattin is with the University of Innsbruck, Technikerstra\ss{}e 21a, 6020 Innsbruck, Austria. E-mail: \url{andrea.burattin@uibk.ac.at}.}}
\date{}

\begin{document}
\maketitle

\begin{abstract}
Process mining represents an important field in BPM and data mining research. Recently, it has gained importance also for practitioners: more and more companies are creating business process intelligence solutions. The evaluation of process mining algorithms requires, as any other data mining task, the availability of large amount of real-world data. Despite the increasing availability of such datasets, they are affected by many limitations, in primis the absence of a ``gold standard'' (i.e., the reference model).

This paper extends an approach, already available in the literature, for the generation of random processes. Novelties have been introduced throughout the work and, in particular, they involve the complete support for multiperspective models and logs (i.e., the control-flow perspective is enriched with time and data information) and for online settings (i.e., generation of multiperspective event streams and concept drifts). The proposed new framework is able to almost entirely cover the spectrum of possible scenarios that can be observed in the real-world.
The proposed approach is implemented as a publicly available Java application, with a set of APIs for the programmatic execution of experiments.
\end{abstract}

\newpage
\tableofcontents

\section{Introduction}

Process mining \cite{VanderAalst2011} gained a lot of attention and is now considered an important field of research, bridging data mining and business process modeling/analysis. In particular, the aim of process mining is to extract useful information from business process executions. Under the umbrella of process mining, different activities could be identified. For example, \emph{control-flow discovery} aims at reconstructing the actual process model starting only from the observations of its executions; \emph{conformance checking} tries to discover discrepancies between the expected (i.e., compliant) executions and the actual ones; \emph{enhancement} extends a process model with additional information obtained from the actual observations.

In data mining, the term \textit{gold standard} (or sometimes also referred as \textit{ground truth}) typically indicates the ``correct'' answer to a mining task (i.e., the reference model). For example, in data clustering, the gold standard may represent the right (i.e., the target) assignment of elements to their corresponding clusters.
Many times, referring to a gold standard is fundamental in order to properly evaluate the quality of mining algorithms \cite{Manning2008,Cios2007}. Several concepts, like \textit{precision} and \textit{recall} are actually grounded on this idea. In general, it is possible to identify the concept of gold standard for all mining tasks.

As for all other mining challenges, the evaluation of new algorithms is difficult.
In order to properly assess the quality of mining algorithms, typically, the evaluation of mining algorithm should be base on real world data.
However in the context of business processes, companies are usually reluctant to publicly share their data for analysis purposes. Moreover, detailed information on their running processes (i.e., the reference models) are considered as company assets and, therefore, are kept private.

Since few years, an annual event, called \emph{BPI challenge}\footnote{See \url{http://www.win.tue.nl/bpi/2015/challenge}.}, releases real world event logs. Despite the importance of this data, the logs are not accompanied with their corresponding gold standards. Moreover, they do not provide examples of all possible real world situations: many times, researchers and practitioners would like to test their algorithms and systems against specific conditions and, to this purpose, those event logs may not be enough. Some other tools, described in the literature (Section~\ref{sec:related}) can be used to construct business processes or to simulate existing ones. However, they are very difficult to use and limited in several aspects (e.g., they can only generate process models, or can simulate just the control-flow perspective).

\subsection{Research Challenges}
\label{sec:challenges}

The final aim of this paper is to support researchers and practitioners in developing new algorithms and techniques for process mining and business process intelligence.
Moreover, we put particular emphasis on the online/stream paradigm which, with the advent of the \emph{big data} and \emph{Internet of things}, is rising interest.
To achieve our goal we have to face the general data availability problem which, in our context, could be decomposed into several research challenges:
\begin{description}
	\item[\textsf{C1}\label{c1}] build large repositories of randomly created process models with control-flow and data perspectives;
	\item[\textsf{C2}\label{c2}] obtain realistic (e.g., noisy) multiperspective event logs, which are referring to a model already known (i.e., the gold standard), to test process mining algorithms;
	\item[\textsf{C3}\label{c3}] generate potentially infinite multiperspective streams of events starting from process models. These strems have to simulate realistic scenarios, e.g., they could contain noise, fluctuating event emission rates, and concept drifts.
\end{description}

\nameref{c1} and \nameref{c2} are required in order to test an approach against several different datasets, and avoid overfitting phenomena (i.e., tailoring an approach to perform well on particular data, but lacking in abstraction).

\nameref{c3} is becoming more and more important due to the emerging importance of big data analysis. Big data is typically characterized~\cite{Gartner,Fan2013} by the
\emph{data volume} and \emph{velocity} (a typical way of dealing with such volume and velocity is via unbounded event streams~\cite{Gama2010,Aggarwal2007});
\emph{variety} (for this, we need multiperspective models, not only with the control-flow perspective);
\emph{variability} (this led us to properly simulate concept drifts~\cite{Burattin2015}).

\vspace{1em}
%\subsection{Contribution}

In this paper we propose a series of algorithms which can be used to randomly generate multiperspective process models (Section~\ref{sec:process-models}). These models can easily be simulated in order to generate multiperspective event logs (Section~\ref{sec:simulation}). Moreover, the whole approach is design keeping the simulation of online settings (Section~\ref{sec:stream-simulation}) in mind: it is possible to generate drifts on the processes (i.e., local evolutions) and it is possible to simulate multiperspective event streams (which are also replicating the drifts).

Therefore, the aim of this paper is twofold: on one hand, we aim at describing the extensions made with respect to our previous work \cite{Burattin2010b}, which constitutes \plg. On the other hand, we want to highlight the research challenges that need to be solved in order to create realistic and useful test data.

The new approach is implemented in a standalone Java application (Section~\ref{sec:implementation}) which is also accompanied by a set of APIs, useful for the programmatic definition of custom experiments.

%\begin{figure}
%	\centering
%	\includegraphics[width=.8\textwidth]{img/components-model.pdf}
%	\caption{Components diagram of the proposed approach.}
%	\label{fig:components}
%\end{figure}
%
%The implemented components are reported in Figure~\ref{fig:components}. The most central component is responsible for the representation of a process model. In order to create a process, the user can import a file; randomly generate a new process; or evolve an existing process into a different one. Processes can also be exported to file. Starting from a process model, \plg can simulate it in order to create an event log or an event stream. Both these last components use a noise generator in order to add more realism to the generated data. In general, this approach makes possible to quickly obtain complex data to perform several types of analysis (Section~\ref{sec:case-study}).
%

In summary, this paper extend the work we presented in \cite{Burattin2010b} since we are now able to:
\begin{itemize}
	\item generate random process models with additional data perspective (or import existing ones);
	\item have a detailed control over the data attributes (e.g., by controlling their values via scripts);
	\item have a detailed control over the time perspective (controlled via scripts);
	\item \emph{evolve} a process model, by randomly changing some of its features (e.g., adding/removing/replacing subprocesses);
	\item generate a realistic multiperspective event log, with executions of a process models and noise addition (with probabilities for different noise behaviors);
	\item generate a stream of multiperspective events referring to process models that could change over time with customizable output ratio.
\end{itemize}

\section{Related Work}
\label{sec:related}

The idea of generating process models for evaluating process mining algorithms has already been explored.

In particular, van Hee and Liu, in \cite{VanHee2010}, presented an approach to generate random Petri nets representing processes. Specifically, they suggest to use a top-down approach, based on a step-wise refinement of Workflow nets \cite{Aalst2004b}, to generate all possible process models belonging to a particular class of Workflow network (also called Jackson nets). This approach has been adopted, for example, in the generation of collections of process model with specific features~\cite{Hee2011}.
A similar and related approach has been reported in \cite{Bergmann2008}, where authors propose to generate Petri nets according to a different set of refinement rules.

In both cases, approaches do not address the problem of generating traces from the developed Petri nets. This task, however, has been explored in the past, in particular for the generation of process mining oriented logs \cite{DeMedeiros2005}. The idea is to decorate a Petri net model, using CPN tool\footnote{See \url{http://cpntools.org}.}, in order to log executions into event log files. Although the approach is extremely flexible and grounded on a solid tool, it suffers of usability drawbacks. The most important problem consists of the complexity of whole procedure, which is also particularly error prone; the complexity in managing timestamps; the impossibility to simulate data objects (i.e., multiperspective models) in a proper way; and impossibility to simulate streams.

The work reported in \cite{Burattin2010b}, extended by this work, provides a first possible complete tool for the random generation of process models and their execution logs.

The approach described in this paper (namely, \plg) extends previous works in two substantial ways. On one hand it improves the generation of random models and their logs by adding data and time perspectives: the new version of \plg is capable of generating random data objects and simulate manually defined ones. Moreover, complete and detailed support for activity and trace timing is provided as well.
Secondly, the whole project has been designed with online settings in mind: it is possible to easily (and automatically) generate random new versions of process models, in order to simulate ``concept drifts''. Moreover, processes can be simulated to generate multiperspective (i.e., with data and preserving temporal relations) event streams.

\section{Process Models in \plg}
\label{sec:process-models}

This section presents the internal representation used to handle business processes. The generation of random business process is reported as well.

\subsection{Internal Representation of Business Processes}
\label{sec:process-representation}

\begin{figure*}
	\includegraphics[width=\textwidth]{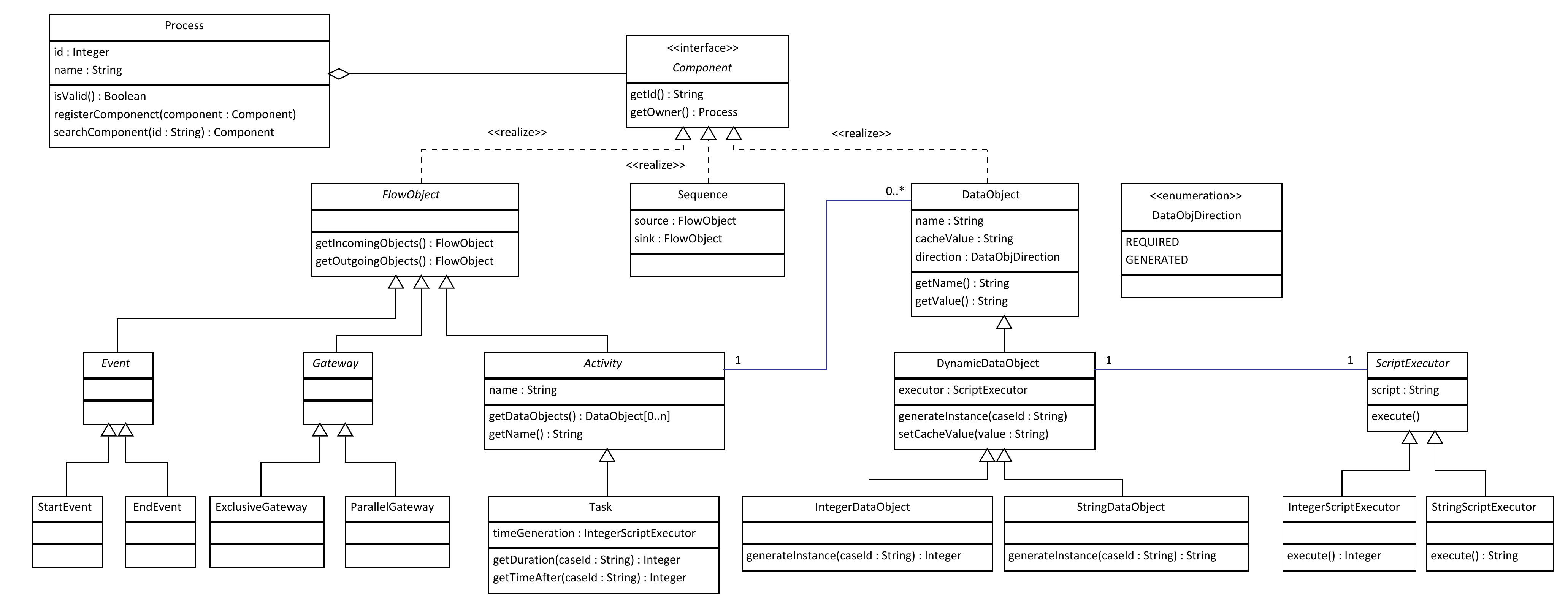}
	\caption{Classes diagram, represented in UML, of the internal structure used for the representation of a process model in \plg. The structure basically reflects a possible instance of a BPMN model diagram.}
	\label{fig:uml-class}
\end{figure*}

In \plg, the internal structure of a process model is actually rather intuitively derived from the definition of a BPMN process model \cite{OMG2009}. Figure~\ref{fig:uml-class} depicts the diagram of the classes involved in the modeling.
In particular, a process is essentially an aggregation of \textit{components}. Each component can be either a flow object, a sequence or a data object. Flow objects are divided into:
\begin{itemize}
	\item events (which are divided into \textit{start} or \textit{end});
	\item gateways (either \textit{exclusive} or \textit{parallel});
	\item tasks (which can be \emph{activities}).
\end{itemize}
Please note that it is not possible to instantiate general events or gateways or tasks (i.e., those classes are abstract). This technique is used to enforce a proper \textit{typing} of such ---otherwise ambiguous--- elements.
Sequences are used to connect two flow objects. A sequence, clearly, imposes a direction of the flow.
Data objects are associated with activities and can be \emph{plain data objects} or \emph{dynamic data objects}. A plain data object is basically a key-value pair. A dynamic data object, instead, has a value that can change every time it is required (i.e., it is dynamically generated by a script). Another characterization of data object is with respect to their ``direction'': a data object can be \textit{generated} or \textit{required} by an activity. These two characterizations play an important role in the process simulation phase. We will get into these details in Section~\ref{sec:simulation}.

\vspace{1em}
With respect to the previous version of this work \cite{Burattin2010b}, we decided to evolve the internal structure into a more general one. This BPMN standard-oriented representation is fundamental in order to allow much more flexibility. For example, now, it is possible to load BPMN models generated with external tools\footnote{An example of supported tool is Signavio, \url{http://www.signavio.com}.}, as long as the modeled components are available also in \plg\ (for example, it is not possible to load a BPMN model with inclusive gateways). However, since we are restricting to non-ambiguous components, we also can convert our processes into more formal languages (e.g., it is possible to convert and export the generated models into Petri nets, using the PNML file format\footnote{See \url{http://www.pnml.org} for more information on this standard.}).

\subsection{Formal Definition of Business Process}
\label{sec:formal-process-definition}

The process representation that we just reported can be structured in a more formal definition. Specifically, a process model $P$ can be seen as a graph $P = (V, E)$, where $V$ is the set of nodes, and $E \subseteq V \times V$ is the set of directed edges. However, since in our context not all nodes or edges are equal, we can improve the definition of $V$ and $E$. Let's then define a process as a tuple $P = ((E_\textit{start}, E_\textit{end}, A, G, D), (S, C))$,  where:
\begin{itemize}
	\item $(E_\textit{start}, E_\textit{end},A,G,D)$ is a tuple, in which each component is a set of nodes with a specific semantic associated. In particular, $E_\textit{start}$ is the set of starting nodes, $E_\textit{end}$ is the set of end nodes, $A$ is the set of activities, $G$ is the set of gateways, $D$ is the set of data objects;
	\item $(S, C)$ is another tuple. Each component of this tuple is a set of edges.\\
	Specifically, $S \subseteq E_\textit{start} \times A \cup A \times E_\textit{end} \cup A \times A \cup A \times G \cup G \times G \cup G \times A$ is a set of sequences connecting process flow objects.\\
	$C \subseteq A \times D \cup D \times A$, such that $\forall d \in D\ |\{(\cdot, d)\in C \}\cup\{(d,\cdot)\in C\}| \leq 1$, is a set of associations going from activities to data objects and from data objects to activities. The additional condition guarantees that one data object is connected with at most one activity.
\end{itemize}

Please note that, the definition just provided partially enforces the semantic correctness of each component involved (for example, it is not possible to connect an end event with a gateway or a data object with an event).

With respect to the data object associations, the $C$ component of a process $P$ permits data objects both incoming and outgoing into and from activities.
This behavior is described in the UML classes diagram, reported in Fig.~\ref{fig:uml-class}, with the \texttt{direction} element of a \texttt{DataObject}.

\section{Random Generation of Business Processes in \plg}
\label{sec:grammar}

The definition of process just described can be used as general representations for the description of relations between activities, events, gateways, and data objects. In this paper, however, we are also interested in the generation of random process models, in order to be able to create a ``process population'' capable of describing several behaviors.

In order to generate random processes, we need to combine some well known workflow control-flow patters \cite{Russell2006,VanderAalst2003}. The patters we are interested in are reported in this summarized list:
\begin{itemize}
	\item \textit{sequence} (WCP-1): direct succession of two activities (i.e., an activity is \textit{enabled} after the completion of the preceding);
	\item \textit{parallel split} (WCP-2): parallel execution (i.e., once the work reaches the split, it is forked into parallel branches, each executing concurrently);
	\item \textit{synchronization} (WCP-3): synchronization of parallel branches (i.e., the work is allowed to continue only when all incoming branches are completed);
	\item \textit{exclusive choice} (WCP-4): mutual execution (i.e., once the work reaches the split, only precisely one of the outgoing branches is allowed to continue);
	\item \textit{simple merge} (WCP-5): convergence of branches (i.e., each incoming branch results in continuing the work);
	\item \textit{structured loop} (WCP-21): ability to execute sub-processes repeatedly.
\end{itemize}
Clearly, these patterns do not describe all the possible behaviors that can be modeled in reality, however we think that most realistic processes are based on them. Actually, we are also going to extended these patters (with the addition of data-objects), in order to generate multiperspective models.

The way we use these patterns is by progressively combining them in order to build a complete process. The combination of such patterns is performed according to a predefined set of rules. We implement this idea via a Context-Free Grammar (CFG) \cite{Hopcroft2006} whose productions are related with the patterns mentioned above. Specifically, we defined the following context-free grammar $G_{\textit{Process}} = \{ V, \Sigma, R, P \}$
where 
$V = \{P, G, G', G_{\LOOP},$ $G_{\AND}, G_{\XOR}, A, A_\textit{act}, A_\textit{do}, D \}$ is the set of the non-terminal symbols,
$\Sigma = \{ \SEQ, (, ), \LOOP, \XOR, \AND, \nearrow, \swarrow, e_\textit{start}, e_\textit{end},$ $a, b, c,$ $\dots,$ $d_1, d_2, d_3, \dots \}$ is the set of all terminals (their ``interpretation'' is described in Table~\ref{tbl:terminal}), $R$ is the set of productions:
$$
	\begin{array}{rcl}
		P  &\rightarrow& e_\textit{start}\ \SEQ\ G\ \SEQ\ e_\textit{end} \\
		G &\rightarrow& G' \mid G_\circlearrowright \\
		G' &\rightarrow& A \mid (G \SEQ G) \mid (A \SEQ G_{\AND} \SEQ A) \mid (A \SEQ G_{\XOR} \SEQ A) \mid \epsilon \\
		G_{\AND} &\rightarrow& G \AND G \mid G \AND G_{\AND} \\
		G_{\XOR} &\rightarrow& G \XOR G \mid G \XOR G_{\XOR} \\
		G_{\LOOP} &\rightarrow& (G' \LOOP G) \\
		A  &\rightarrow& A_\textit{act} \mid A_\textit{do} \\
		A_\textit{do} &\rightarrow& \DOIn{A_\textit{act}}{D} \mid \DOOut{A_\textit{act}}{D} \\
		A_\textit{act} &\rightarrow& a \mid b \mid c \mid \dots \\
		D &\rightarrow& d_1 \mid d_2 \mid d_3 \mid \dots
	\end{array}
$$
and $P$ is the starting symbol for the grammar.
\begin{table}%[t]
	\centering
	\begin{tabular}{cp{25em}}
		\toprule
		\textbf{Symbols} & \multicolumn{1}{c}{\textbf{Meaning}} \\
		\midrule
		$e_\textit{start}$ & The start event of the process \\
		$e_\textit{end}$ & The end event of the process \\
		( ) & Parentheses are used to describe operators precedence \\
		$\SEQ$ & Operator, it indicates a sequential connection \\
		$\LOOP$ & Operator, repetition of the first parameter by executing the second \\
		$\AND$ & Operator, its parameters executed in parallel (``AND'') \\
		$\XOR$ & Operator, its parameters executed in mutual exclusion (``XOR'') \\
		$\DOIn{a}{d}$ & Indicates that data object $d$ is required by activity $a$ \\
		$\DOOut{a}{d}$ & Indicates that data object $d$ is generated by activity $a$ \\
		$a$, $b$, $c$, \dots & The set of possible activity names \\
		$d_1$, $d_2$, $d_3$, \dots & A set of possible data objects \\
		\bottomrule
	\end{tabular}
	\caption{All the terminal symbols of the context-free grammar used for the random generation of business processes and their corresponding meanings.}
	\label{tbl:terminal}
\end{table}
Using this grammar, a process is described by a string derived from $G_{\textit{Process}}$.

Analyzing the production rules, it is possible to see that each process requires a starting and a finishing event and, in the middle, there must be a sub-graph $G$. A sub-graph can be either a ``simple sub-graph'' ($G'$) or a ``repetition of a sub-graph'' ($G_{\LOOP}$).

Starting from the first case: a sub-graph $G'$ can be a single activity $A$; the sequential execution of two sub-graphs $(G;G)$; the exclusive or parallel execution of some sub-graphs (respectively, $(A; G_{\XOR}; A)$ and $(A; G_{\AND}; A)$); or an ``empty'' sub-graph $\epsilon$.
It is important to note that the generation of parallel and mutual exclusion branches is always ``well structured''.

Analyzing the repetition of a sub-graph ($G_{\LOOP}$) it should be noticed that, semantically, the repetition of a sub-graph $(G' \LOOP G)$ is described as follows: each time we want to repeat the ``main'' sub-graph $G'$, we have to perform another sub-graph $G$; the idea is that $G$  (that can even be only a single or empty activity) corresponds to the ``roll-back'' activities required in order to prepare the system to the repetition of $G'$ (which, also, could be a empty activity).

The structure of $G_{\AND}$ and $G_{\XOR}$ is simple and expresses the parallel execution or the choice between at least 2 sub-graphs.

$A$ represents the set of possible activities. In this case two productions are possible: $A_\textit{act}$ which generates just an activity, or $A_\textit{do}$ which generates an activity with a data object associated. In this latter case, two more productions are possible: $\DOIn{A_\textit{act}}{D}$ and $\DOOut{A_\textit{act}}{D}$: the first generates an activity with a required data object, the second produces an activity with a generated data object.
Finally, the grammar defines activities just as alphabetic identifiers but, actually, the implemented tool ``decorates'' it with other attributes, such as a unique identifier. The same observation holds for data objects.

Finally, this grammar definition allows for more activities with the same name, however in our implemented generator all the activities are considered to be different.

\begin{figure}
	\centering
	\subfloat[Example of derivation tree. Note that, for space reason, we omitted the explicit representation of some basic productions.] {
	\begin{tikzpicture}
		[
			level distance=.8cm,
			nont/.style={rectangle, draw=gray, rounded corners=1mm, fill=gray!10, text centered, anchor=north, text=black},
			term/.style={rectangle, draw=black, rounded corners=1mm, fill=gray!30, text centered, anchor=north, text=black}
		]
		\node [nont] {$P$}
			child {node [nont] {$e_\textit{start} \SEQ G \SEQ e_\textit{end}$}
			child {node[term] {$e_\textit{start}$}}
			child {node [nont] {$G$}
				child {[sibling distance=6cm] node [nont] {$(G;G)$}
					child { node [nont] {$A$}
						child {node [nont] {$A_\textit{act}$}
							child {node [term] {$a$}}
						}
					}
					child {[sibling distance=3cm] node [nont] {$(G \SEQ G)$}
						child {node [nont] {$(G' \LOOP G)$}
							child {[sibling distance=1.3cm] node [nont] {$A;(G \AND G);A$}
								child {node [nont] {$A_\textit{act}$}
									child {node [term] {$b$}}
								}
								child {node [nont] {$A_\textit{do}$}
									child {node [nont] {$\DOOut{A_\textit{act}}{D}$}
										child {node [term] {$\DOOut{c}{d_1}$}}
									}
								}
								child {node [nont] {$A$}
									child {node [nont] {$A_\textit{act}$}
										child {node [term] {$d$}}
									}
								}
								child {node [nont] {$A_\textit{act}$}
									child {node [term] {$e$}}
								}
							}
							child {node [nont] {$A$}
								child {node [nont] {$A_\textit{act}$}
									child {node [term] {$f$}}
								}
							}
						}
						child {node [nont] {$A$}
							child {node [nont] {$A_\textit{act}$}
								child {node [term] {$g$}}
							}
						}
					}
				}
			}
			child {node[term] {$e_\textit{end}$}}
			};
	\end{tikzpicture}
	}
	
	\subfloat[The string derived from the above tree.]{
		$e_\textit{start} \SEQ \left(a \SEQ \left(\left(b \SEQ \left(\DOOut{c}{d_1} \AND{} d \right) \SEQ e \LOOP f\right) \SEQ g\right)\right) \SEQ e_\textit{end}$
	}
	
	\subfloat[BPMN representation, created by \plg{}, for the process generated.]{
		\includegraphics[width=\textwidth]{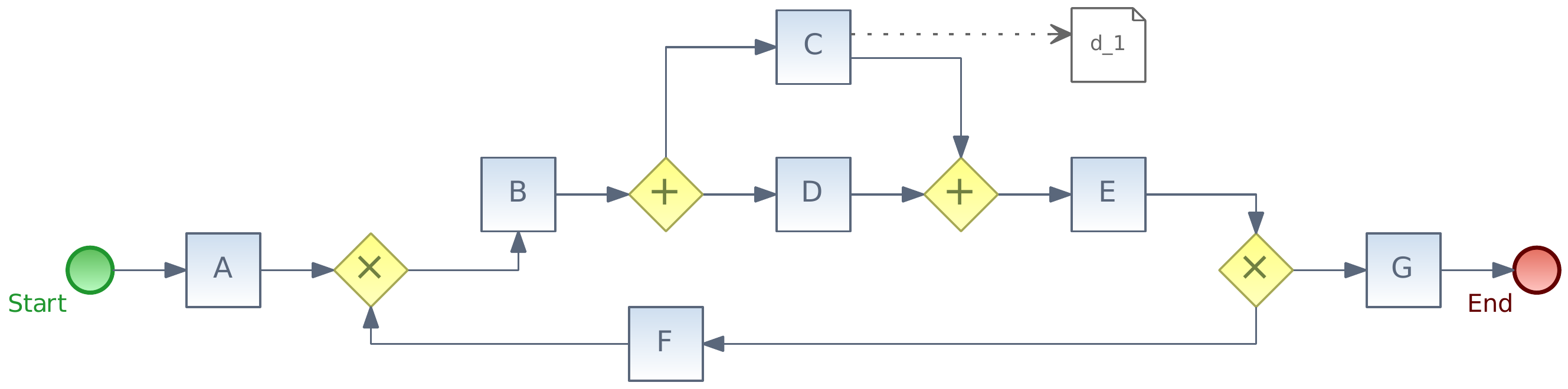}
	}
	
%	\subfloat[Petri net representation, generated by \plg{}, for the process generated (only the control-flow perspective is reported in this case).]{
%		\includegraphics[width=.47\textwidth]{img/derivation-example/petri-derivation-example.pdf}
%	}
	\caption{Derivation tree of a process and its string and BPMN representation.}
	\label{fig:derivation-example}
\end{figure}
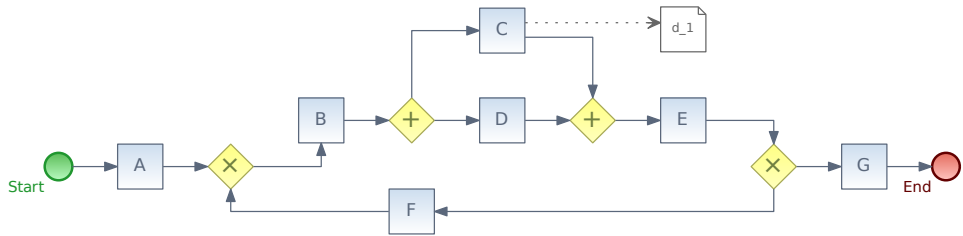

In Fig.~\ref{fig:derivation-example} an example of all the steps involved in the generation of a process
are shown: the derivation tree, the string of terminals, and two graphical representations of the final process (using BPMN and Petri net notations).

\subsection{Grammar Capabilities}

The context free grammar just provided is not capable of generating all the possible business models that could be described using languages such BPMN or Petri net. In particular, we are restricting to block structured ones \cite{Kopp2009}.
Although restricting to block structured processes might seems rough, these processes benefit from very interesting properties \cite{Vanhatalo2008}. Moreover, recently, the process mining community started to focus on this types of processes \cite{Buijs2012,VanderAalst2012a,Leemans2013}, especially for the soundness properties that they can guaranteed.
With the adoption of the context-free grammar proposed, we decided to stick to this type of language as well.

Please note that the block structure restriction only affects the random process generation part of \plg: all other components (i.e., process evolution, and simulations for generation of event logs or stream) are still functioning also with imported (and non-block structured) processes.

We can also note that there is a straightforward translation of a string produced by the \plg grammar into the graph representation introduced in the previous sections. Therefore, the processes generated with \plg can always be expresses as BPMN (with all perspectives) or Petri net (with just the control-flow perspective).

\subsection{Grammar Extension with Probabilities}

As previously stated, we want to randomly generate strings of terminal using the context-free grammar described earlier. However, in order to provide the user with a deep control over the final structure of the generated processes, we converted the CFG into a stochastic context-free grammar (SCFG) \cite{Chomsky1956,Chomsky1963}. This type of grammars have also been widely used for modeling RNA structures \cite{Sukosd2014,Sakakibara1994}.

Specifically, to adopt this models, we need to add probabilities associated to each production rule. This allows us to introduce user-defined parameters to control the presence of specific pattern into the generated process. These are the probabilities defined (with indication on whether the user is asked to provide the value):
$$
	\begin{array}{lcrcll}
		\pi_1 & \text{\ for\ } & G & \rightarrow & G_{\LOOP} & \text{required} \\
		\pi_2 & \text{\ for\ } & G & \rightarrow & G' & \text{as } 1 - \pi_1 \\[1em]
		\pi_3 & \text{\ for\ } & G' & \rightarrow & A & \text{required} \\
		\pi_4 & \text{\ for\ } & G' & \rightarrow & (G;G) & \text{required} \\
		\pi_5 & \text{\ for\ } & G' & \rightarrow & (A;G_{\AND};A) & \text{required} \\
		\pi_6 & \text{\ for\ } & G' & \rightarrow & (A;G_{\XOR};A) & \text{required} \\
		\pi_7 & \text{\ for\ } & G' & \rightarrow & \epsilon & \text{required} \\[1em]
		\pi_8 & \text{\ for\ } & G_{\AND} & \rightarrow & G \AND G_{\AND} & \text{computed} \\
		\pi_9 & \text{\ for\ } & G_{\AND} & \rightarrow & G \AND G & \text{as } 1 - \pi_8 \\[1em]
		\pi_{10} & \text{\ for\ } & G_{\XOR} & \rightarrow & G \XOR G_{\XOR} & \text{computed} \\
		\pi_{11} & \text{\ for\ } & G_{\XOR} & \rightarrow & G \XOR G & \text{as } 1 - \pi_{10} \\[1em]
		\pi_{12} & \text{\ for\ } & A & \rightarrow & A_\textit{do} & \text{required} \\
		\pi_{13} & \text{\ for\ } & A & \rightarrow & A_\textit{act} & \text{as } 1 - \pi_{12} \\
	\end{array}
$$

In order to have a valid grammar the system has to enforce that the probabilities of each production sum to 1. Let's define the groups probabilities as: $G_\textit{Pr} = \{ \{\pi_1, \pi_2\}, \{\pi_3, \dots, \pi_7\}, \{\pi_8, \pi_9\}, \{\pi_{10}, \pi_{11}\},$ $\{\pi_{12}, \pi_{13}\} \}$. Then the following property has to be fulfilled:
$
	\forall\ \textit{Pr} \in G_\textit{Pr} \quad \sum_{p \in \textit{Pr}} p = 1
$.
As you can see, this property holds by construction for $\{\pi_1, \pi_2\}$ and for $\{\pi_{12}, \pi_{13}\}$. For $\{\pi_3, \dots, \pi_7\}$ it is artificially enforced (the user is required to insert weights, which are then proportionally adapted in order to sum up to 1).

The two remaining sets (i.e., $\{\pi_8, \pi_9\}$ and $\{\pi_{10}, \pi_{11}\}$) are treated slightly differently: in this case the user is required to insert the maximum number of possible AND/XOR branches. This information let us dynamically compute the probability values. Let's consider the AND case: in the beginning, $\pi_8=\pi_9=0.5$. The system keeps these values unchanged until the maximum number of AND branches are generated (i.e., the number of times that the production rule $G_{\XOR} \rightarrow G \XOR G_{\XOR}$ is consecutively executed). Once the max value is reached, probabilities are changed in order to stop generating more branches: $\pi_8=0$ (and therefore $\pi_9=1$). Similar approach is adopted for the XOR branches (i.e., $\{\pi_{10}, \pi_{11}\}$). Although this adaptation forces the context-free property of the grammar, we think that, for the final user, it is much more easy to specify the maximum number of branches instead of the actual probabilities.

In order to provide the user with a more detailed control of the grammar, we require to specify an additional parameter, which is called \textit{maximum depth}. This parameter allows the user to control the depth of the derivation tree: once the tree reaches the \textit{maximum depth},  probabilities are artificially changed to these values: $\pi_3=\pi_7=0.5$, $\pi_4=\dots=\pi_6=\pi_8=\pi_{10}=0$. This probabilities change forces the derivation tree to limit its depth by allowing only new activity or skip patters.

\section{Process Simulation in \plg}
\label{sec:simulation}

In order to evaluate process mining algorithms or, in general, to stress business intelligence systems, we are not only interested in the random generation of processes, but we also need observations of the activities executed for each process instance, i.e. event logs. This section reports details on how we generate multiperspective logs and how to make them more realistic by artificially inserting some noise.

Before getting into the actual simulation algorithms, it is important to define the concept of event log. In order to better understand how an event log is composed, we clarify that an execution of a business process forms a \emph{case}. The sequence of events in a case is called \emph{trace}, and each trace, in turn, consists of the list of \emph{events} which refer to specific activities performed. It is possible to see each event as a set of attributes (i.e., key-value pairs). The fundamental attributes of an event are:
\textit{(i)} the \textit{name} of the executed activity,
\textit{(ii)} the \textit{timestamp} (which reports the execution time of the given event) and
\textit{(iii)} the activity \textit{lifecycle} (whether the event refers to the \textit{beginning} or to the \textit{completion} of an activity). The lifecycle attribute is important when a recorded activity lasted for a certain amount of time (i.e., it is not instantaneous): in this case, two events are recorded, one when the activity begins and another when the activity ends.

More formally, given the set of all possible activity names $\cal{A}$, the set of all possible case identifiers $\cal{C}$, the set of timestamps $\cal{T}$, and the set of lifecycle transitions $\cal{L} =\{\textit{start},\textit{complete}\}$, it is possible to define an \emph{event} $e$ as a tuple, such as $e = (c,a,t,l) \in \cal{C} \times \cal{A} \times \cal{T} \times \cal{L}$. In this case, it describes the occurrence of activity $a$, with lifecycle transition $l$, for the case $c$, at time $t$. Please note that the attributes reported here are just the minimum required ones: other attributes can be added to the event (in general, each data object of the process will generate a new attribute).
Given an event $e = (c, a, t, l, a_1, \dots, a_k)$, it is possible to extract each field using a projection operator: $\#_\text{case}(e) = c$; $\#_\text{activity}(e) = a$; $\#_\text{time}(e) = t$; $\#_\textit{lifecycle}(e) = l$, and so on.

Given a finite set $\seq{n} = \{1, 2, \dots, n\}$ and a ``target'' set $A$, we define a sequence $\sigma$ as a function $\sigma: \seq{n} \to A$. We say that $\sigma$ maps indexes to the corresponding elements in $A$. For simplicity, we refer to a sequence using its ``string'' interpretation: $\sigma = \langle s_1, \dots, s_n \rangle$, where $s_i = \sigma(i)$ and $s_i \in A$. Moreover, we assume to have concatenation and cardinality operators: respectively $\left\langle e^1_1, \dots, e^1_n \right\rangle \cdot \left\langle e^2_1, \dots, e^2_m \right\rangle = \left\langle e^1_1, \dots, e^1_n, e^2_1, \dots, e^2_m \right\rangle$ and $\left|\left\langle e_1, \dots, e_n \right\rangle\right| = n$.

In our context, we use timestamps to sort the events. Therefore, it is safe to consider a trace just as a sequence of events. In turn, a log is just a set of traces. Therefore, traces are allowed to overlap: given a log $l$ with two traces $t_1 = \left\langle e^1_1, \dots, e^1_n \right\rangle \in l$ and $t_2 = \left\langle e^2_1, \dots, e^2_m \right\rangle \in l$ it is possible to have that $\#_\textit{time}\left(e^1_1\right) \leq \#_\textit{time}\left(e^2_1\right) \leq \#_\textit{time}\left(e^1_n\right)$ or $\#_\textit{time}\left(e^2_1\right) \leq \#_\textit{time}\left(e^1_1\right) \leq \#_\textit{time}\left(e^2_m\right)$.

\subsection{Multi-Perspective Simulation}

The procedure for the generation of logs out of business process, basically, consists of a simulation engine running a ``\textit{plain-out activity}'' \cite{VanderAalst2011}. However, in order to properly simulate all the perspectives required, some conventions need to be defined.

The structure of process models that \plg can handle is restricted to the family of BPMN models with an unambiguous semantic. Therefore, in \plg, it is possible to consider a process as its equivalent Petri net representation \cite{Peterson1977,Murata1989}. The main advantage, in this case, is that it is possible to play the token-game for simulating the process.

The procedures for the simulation of a process instance are reported in Algorithm~\ref{alg:Simulate}, \ref{alg:SimulateProcess} and \ref{alg:SimulateActivity}. These procedures use the following additional functions: $\incoming$, $\out$ and $\random$. Let's assume a process $P = ((E_\textit{start}, E_\textit{end}, A, G, D),$ $(S, C))$, as described in Section~\ref{sec:formal-process-definition}. Given $c\in A \cup G$, we can define $\incoming(c) = \{ c' \mid (c', c) \in S \}$ and $\out(c) = \{ c' \mid (c, c') \in S \}$. $\random(s)$, instead, given a general set $s$, returns a randomly selected element $e$ such that $e \in s$.

Algorithm~\ref{alg:Simulate} represents the main entry point of the simulation: it expects, as input, a process model and the number of traces to simulate. Then, it basically iterates the generation of single traces in order to populate the log. Line~\ref{alg:Simulate:sort} is required in order to properly sort the events, and line~\ref{alg:Simulate:noise} introduces, if required, some noise into the trace. The noise generation will be described in Section~\ref{sec:noise-addition}.

\begin{algorithm}
	\caption{A general simulation procedure \label{alg:Simulate}}
	\LinesNumbered
	\DontPrintSemicolon
	\Input{%
		$P = ((E_\textit{start}, E_\textit{end}, A, G, D), (S, C))$: the process to simulate \\
		$\textit{tot}$: the number of traces to generate
	}
	\Output {An event log}
	log $\gets \emptyset$ \;
	\For{$i =1$ \emph{\textbf{up to}} $\textit{tot}$} {
		$t \gets \langle\ \rangle$ \Comment*{Generate a new trace}
		SimulateProcess($P$, $t$, $\random(E_\textit{start})$, $\bot$)$\}$ \Comment*{Algorith \ref{alg:SimulateProcess}}
		\lnl{alg:Simulate:sort} sort($t$) \Comment*{Sort events w.r.t. their times}
		\lnl{alg:Simulate:noise} add trace-level noise to $t$ \Comment*{See Section~\ref{sec:noise-addition}}
		log $\gets \text{log} \cup \{t\}$ \Comment*{Add the trace to the log}
	}
	\Return log \;
\end{algorithm}

Algorithm~\ref{alg:SimulateProcess} is in charge of the control-flow simulation. The algorithm expects as input the process to simulate, the component to analyze, and the sequence (i.e., the edge) that brought the analysis to the current component. First of all, the algorithm requires the definition of a global set $t$. This set is fundamental for the ``token game'': making an analogy with Petri nets, it stores the current marking (i.e., the tokens configuration). In our case, however, the set contains the edges that are ``allowed to execute''.

\begin{algorithm}[!htb]
	\caption{Simulate Process \label{alg:SimulateProcess}}
	\footnotesize
	\LinesNumbered
	\DontPrintSemicolon
	
	\Input{%
		$P = ((E_\textit{start}, E_\textit{end}, A, G, D), (S, C))$: the process to simulate\\
		$t$: the trace containing the simulated events\\
		$c$: process component to simulate\\
		$s = (i, c)$: incoming sequence
	}
	\BlankLine
	$\text{tokens} \gets $ globally defined set of tokens (i.e., sequences), initially the empty set \;
	\BlankLine
	\uIf{$c$ \emph{is a Task} \emph{\textbf{or}} $c$ \emph{is a XOR gateway}} {
		\If{$c$ \emph{is a Task}} {
			Simulate Activity($P$, $t$, $c$) \Comment*{Algorithm~\ref{alg:SimulateActivity}} \label{alg:SimulateProcess:act}
			$\text{tokens} \gets \text{tokens} \setminus \{s\}$ \;
		}
		\If{$|\out(c)| \geq 1$} {
			$n \gets \random(\out(c))$ \Comment*{Randomly select the following component}
			$\text{tokens} \gets \text{tokens} \cup \{(c,n)\}$ \Comment*{Update tokens}
			Simulate Process($P$, $t$, $n$, $(c, n)$) \Comment*{Recursion} \label{alg:SimulateProcess:actRec}
		}
	}
	\ElseIf{$c$ \emph{is an AND gateway}} {
		\eIf(\Comment*[f]{We treat $c$ as a split}){$|\out(c)| > 1$} { \label{alg:SimulateProcess:split}
			$\text{tokens} \gets \text{tokens} \setminus \{s\}$ \;
			\ForAll(\Comment*[f]{Add all tokens}){$n \in \out(c)$} {
				$\text{tokens} \gets \text{tokens} \cup \{(c, n)\}$ \;
			}
			\ForAll{$n \in \out(c)$} {
				Simulate Process($P$, $t$, $n,(c, n)$) \Comment*{Recursive call} \label{alg:SimulateProcess:split-call}
			}
		}(\Comment*[f]{In this case, we treat $c$ as a join}){ \label{alg:SimulateProcess:join}
		allBranchesSeen $\gets$ \textbf{true} \; \label{alg:SimulateProcess:check-start}
		\Comment{Check whether all branches (i.e., incoming edges) have been executed}
		\ForAll{$p \in \incoming(c)$} {
			\If{$(p, c) \notin t$} {
				allBranchesSeen $\gets$ \textbf{false} \;
				break \;
			}
		} \label{alg:SimulateProcess:check-end}
		\If{\emph{allBranchesSeen is \textbf{true}}} {
			\ForAll(\Comment*[f]{Remove tokens}){$p \in \incoming(c)$} {
				$\text{tokens} \gets \text{tokens} \setminus \{(p,c)\}$ \;
			}
			$n \gets \out(c)$ \Comment*{Get the outgoing edge}
			$\text{tokens} \gets \text{tokens} \cup \{(c,n)\}$ \Comment*{Update tokens}
			Simulate Process($P$, $t$, $n$, $(c, n)$) \Comment*{Recursive call} \label{alg:SimulateProcess:and-continue}
		}
	}
}
\end{algorithm}

The idea behind Algorithm~\ref{alg:SimulateProcess} is to call itself on all elements (events, tasks, and gateways) of the process. Then different behaviors are performed, based on the analyzed element.
Specifically, if the element is a task, it is simulated (line~\ref{alg:SimulateProcess:act}), and then the algorithm is called on the following component (line~\ref{alg:SimulateProcess:actRec}).
If the analyzed element is a XOR gateway, then the call is just passed to one (randomly picked) outgoing element (line~\ref{alg:SimulateProcess:actRec}).
If the currently analyzed element is an AND gateway, we made the assumption that it can be either a split or a join (not both at the same time). It is possible to discriminate between split and join by checking the number of outgoing edges (line~\ref{alg:SimulateProcess:split} and \ref{alg:SimulateProcess:join}). If the gateway is an AND split, it is necessary to make one call for each AND branch (line~\ref{alg:SimulateProcess:split-call}). If the gateway is a join, then it is necessary to check whether all the incoming branches are terminated (lines~\ref{alg:SimulateProcess:check-start}-\ref{alg:SimulateProcess:check-end}). If this is the case, then the flow is allowed to continue with the following activities (line~\ref{alg:SimulateProcess:and-continue}).
Please note that we omitted here the description of the token handling (i.e., insertion, check, and removal) for readability purposes: it is managed in a standard way.

Algorithm~\ref{alg:SimulateActivity} is responsible for adding an activity to the provided trace. The algorithm first creates the \emph{start} event for the activity and populates it with the standard fields (lines~\ref{alg:SimulateActivity:createStart}-\ref{alg:SimulateActivity:lifeStart}). If the activity has a non-instantaneous duration, the algorithm also creates a \emph{complete} event (line~\ref{alg:SimulateActivity:createComplete}-\ref{alg:SimulateActivity:lifeComplete}).
\begin{algorithm}[h!]
	\caption{Simulate Activity \label{alg:SimulateActivity}}
	\LinesNumbered
	\DontPrintSemicolon

	\Input{%
		$P = ((E_\textit{start}, E_\textit{end}, A, G, D), (S, C))$: the process \\
		$t$: the trace that contain the new events \\
		$a$: activity to simulate
	}
	\Comment{Generate the activity start event}
	$e_\textit{start} \gets $ new event referring to activity $a$ \; \label{alg:SimulateActivity:createStart}
	$\#_\textit{activity}(e_\textit{start}) \gets $ the name of activity $a$ \;
	$\#_\textit{time}(e_\textit{start}) \gets $ activity time \Comment*{Details in text}
	$\#_\textit{lifecycle}(e_\textit{start}) \gets \emph{start}$ \; \label{alg:SimulateActivity:lifeStart}
	\Comment{Decorate with all generated data objects}
	\ForAll{$d \in \{d \mid (a, d) \in C\}$} {
		$\#_\textit{d}(e_\textit{start}) \gets $ value generated for $d$ \;
	}
	\Comment{Decorate with all required data objects}
	\If{$|t| > 1$} {
		\ForAll{$d \in \{d \mid (d, a) \in C\}$} {
			lastEvent $\gets t(|t| - 1)$ \;
			$\#_\textit{d}($lastEvent$) \gets $ value generated for $d$ \;
		}
	}
	add event-level noise to $e_\textit{start}$ \Comment*{See Section~\ref{sec:noise-addition}}
	$t \gets t \cdot \langle e_\textit{start} \rangle$ \;
%	\BlankLine
	\Comment{Generate the activity completion event}
	\If{activity $a$ is not instantaneous} {
		$e_\textit{complete} \gets $ new event referring to activity $a$ \; \label{alg:SimulateActivity:createComplete}
		$\#_\textit{activity}(e_\textit{complete}) \gets $ the name of activity $a$ \;
		$\#_\textit{time}(e_\textit{complete}) \gets \#_\textit{time}(e_\textit{start}) + $ activity duration \;
		$\#_\textit{lifecycle}(e_\textit{complete}) \gets \emph{complete}$ \; \label{alg:SimulateActivity:lifeComplete}
		\Comment{Decorate with all generated data objects}
		\ForAll{$d \in \{d \mid (a, d) \in C\}$} {
			$\#_\textit{d}(e_\textit{start}) \gets $ value generated for $d$ \;
		}
		add event-level noise to $e_\textit{complete}$ \Comment*{See Section~\ref{sec:noise-addition}}
		$t \gets t \cdot \langle e_\textit{complete} \rangle$ \;
	}
\end{algorithm}

In order to determine the activity time and its duration, the system needs to check whether the user specified any of these parameters. If no specifications are reported, then the activity is assumed to be instantaneous and to execute a fixed amount of time after the previous one. However, as said, the user can manually specify these parameters. To do so, the user has to provide two Python \cite{python} functions: \texttt{time\_after(caseId)} and \texttt{time\_lasted(caseId)}.
Both these functions are called by the simulator with the \texttt{caseId} parameter valued to the actual case id: this allows the two functions to be case-dependent (for example, it is possible to save files with contextual information).
Specifically, \texttt{time\_} \texttt{after(caseId)} is required to return the number of seconds that have be (virtually) waited before the following activity is allowed to start. \texttt{time\_lasted(caseId)}, instead, has to return the number of seconds that the activity is supposed to last.
This approach is extremely flexible, and allows the user to make very complex simulations. For example, it is possible to define different durations for the same activity depending on which flow the current trace has followed so far, or with respect to the number of iterations on a loop. Examples of such functions are reported in listing~\ref{lst:time-functions}.
\begin{lstlisting}[float,style=customPython, caption={Example of random activity duration (between 5 and 15 minutes) and random time after the execution of an activity (between 1 and 5 minutes).}, label=lst:time-functions]
from random import randint
# This Python script is called for the generation of the time related features of the activity. Note that the functions parameters are the actual case id of the ongoing simulation (you can use this value for customize the behavior according to the actual instance).

# The time_after(caseid) function is returns the number of second to wait before the following activity can start.
def time_after(caseid):
    return randint(60*1, 60*5)

# The time_lasted(caseid) function returns the number of seconds the activity is supposed to last
def time_lasted(caseid):
    return randint(60*5, 60*15)
\end{lstlisting}

Once the time-related properties of an activity are computed, Algorithm~\ref{alg:SimulateActivity}, has to deal with the data objects associated with the current activity. In particular, \emph{generated} data objects (see Section~\ref{sec:process-representation}) are supposed to generate values written as the current activity's attribute. \emph{Required} data objects, instead, are written as attributes for the activity which precedes the current one in the trace. The ratio behind this decision is that \emph{generated} data objects are assumed as values written as output of the current activity. \emph{Required} data objects, instead, are variables that has to be observed prior to the execution of the current activity. However, since the simulation is driven by the control-flow, it is necessary to adjust the variable values \emph{a posteriori}.

\begin{figure}
	\centering
	\includegraphics[width=.4\textwidth]{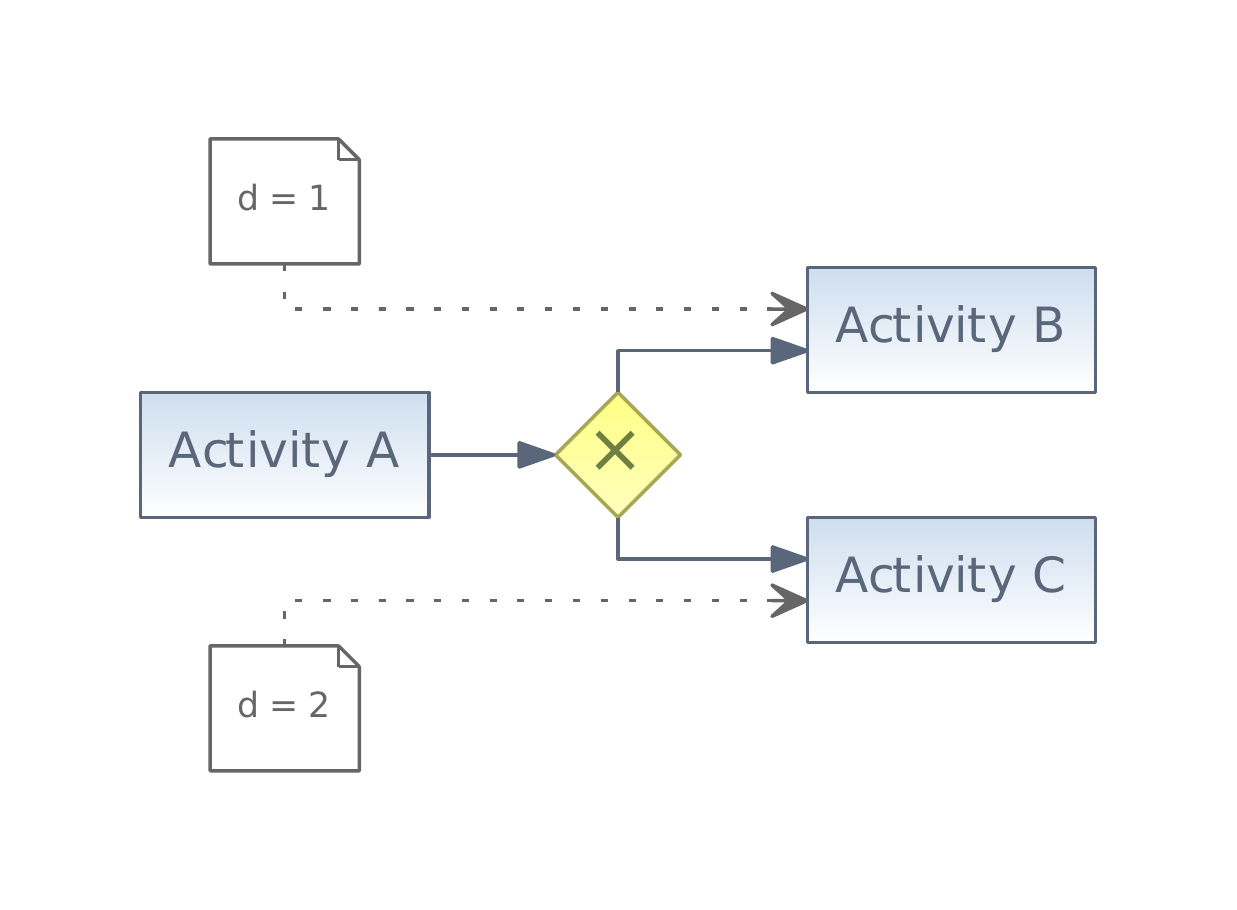}
	\caption{A process fragment of a XOR split gateway with two branches, each of them starting with a different required data object.}
	\label{fig:useful-required-do}
\end{figure}

In order to better understand the utility of \emph{required} data objects, let's consider the process fragment reported in Figure~\ref{fig:useful-required-do}. In this case, the simulation will first perform ``Activity A'' and then either ``Activity B'' or ``Activity C''. However, all the times the simulation engine generates ``Activity B'', it also decorates the event referring to ``Activity A'' (belonging to the same trace) with \texttt{d = 1}. Instead, all the times the simulation engine generates ``Activity C'', it also decorates the event referring to ``Activity A'' (belonging to the same trace) with \texttt{d = 2}. Therefore, an analysis system fed with such example trace could infer a correlation between the value of the attribute \texttt{d} of ``Activity A'', and the following activity.

From the characterization reported in Fig.~\ref{fig:uml-class}, and described in Section~\ref{sec:process-models}, it is possible to distinguish two types of data objects: \emph{plain data objects} and \emph{dynamic data objects}. This distinction is required by the simulation engine in order to properly deal with them: plain data objects are treated as fixed values (i.e., the simulation generates always the same value); dynamic data objects are actually Python scripts whose values are determined by the execution of the script itself. These scripts must implement a \texttt{generate(caseId)} function which is supposed to return either an integer or a string value (depending on the type of data object). An example of integer dynamic data object script is reported in listing~\ref{lst:generate-do}.
\begin{lstlisting}[float,style=customPython, caption={Example of script for the generation of random integer values (in the range 0, 1000).}, label=lst:generate-do]
from random import randint
# This Python script is called for the generation of the integer data object. Note that the parameter of this function is the actual case id of the ongoing simulation (you can use this value to customize your data object). The function name has to be "generate".

def generate(caseId):
    return randint(0, 1000)
\end{lstlisting}
Please note that, also in this case, the function is called with the \texttt{caseId} parameters valued with the actual instance's case id, providing the user with an in-depth, and case dependent, control over the generated values. \texttt{ScriptExecutor} component (and its subclasses), reported in Figure~\ref{fig:uml-class}, are in charge of the execution of the Python scripts.

There is no particular limit on the number of plain and dynamic data objects that a task can have, both required and generated. Clearly, the higher the number of data objects to generate, the longer the simulation will take.

The current random process generator component is only able to generate plain data objects. Specifically, the generated data objects are named \texttt{variable\_a}, \texttt{variable\_b}, \dots\ and the values they return are just random strings.

Considering all the simulation aspects described in this section we can conclude that our approach is able to simulate multiperspective models in order to generate multiperspective logs. The ``multiperspective'' term, in this context, means that the data generate does not only refer to the control-flow perspective, but have also detailed timing properties and the data generate could be extremely articulated and tailored to the actual simulation scenario.

\subsection{Noise Addition}
\label{sec:noise-addition}

In order to generate more realistic data, we introduced a noise component.

The noise component plays a role after the process has been simulated and a trace is available. Specifically, this trace is fed to the noise component which could apply noise at three different ``levels'':
\emph{(i)} at the trace level (i.e., noise which involve the trace organization);
\emph{(ii)} at the event level (i.e., noise which involve events on the control-flow perspective);
\emph{(iii)} at the data object level (i.e., noise which involve the data perspective associated to events).
The actual noise generation is driven by the parameters set by the user. Such parameters, basically, indicate the probability of applying a particular noise type to the trace. Setting all these values to zero implies having trace with no noise.

The noise details for the trace and -partially- for the event level have already been discussed in the literature and reported in details in \cite{Gunther2009phd,Medeiros2006}. The idea is that the user has to specify the probability of all the different noise events, and the simulator will apply the corresponding effect. Possible trace-level noise phenomena are:
\begin{itemize}
	\item a trace which is missing its \emph{head} (i.e., its first events). In this case the user has also to specify the maximum size for a head (which will be randomly chosen between 1 and the provided value);
	\item a trace which is missing its \emph{tail} (i.e., its last events). In this case the user has also to specify the maximum size for the tail (which will be randomly chosen between 1 and the provided value);
	\item a trace which is missing an \emph{episode} (i.e., a sequence of contiguous events). In this case the user has also to specify the maximum size for an episode (which will be randomly chosen between 1 and the provided value);
	\item an alien event introduced into the trace, in a random position, with random attributes;
	\item a doubled event on the trace.
\end{itemize}
Possible noise effects at the event level are:
\begin{itemize}
	\item the random change of the activity name of an event;
	\item the perturbed order between two events of a trace (since the timestamp attributes of two events are involved, we consider this as an event-level noise).
\end{itemize}
Finally, possible data object-level noises are:
\begin{itemize}
	\item random modification of an integer dynamic data object. In this case the user has also to specify the maximum value $\Delta$ of the change: given the old value $v$, the new one (i.e., after noise) will be $v + \delta$, with $\delta$ random in the closed interval $[-\Delta,+\Delta]$;
	\item random modification of a string dynamic data object (replacement of the current string with a randomly generated new one).
\end{itemize}

In order to simplify the noise configuration, we already defined some basic ``noise profiles'', such as:
\emph{(i)} complete noise;
\emph{(ii)} noise only on the control-flow;
\emph{(iii)} noise only for data-objects;
\emph{(iv)} noise only on activity names;
\emph{(v)} no noise at all.

\section{Stream Simulation in \plg}
\label{sec:stream-simulation}

As stated before, \plg explicitly was design for the simulation of online event streams. Specifically, in this context, we adopted the definition of stream already used in the process mining community \cite{Maggi2013,Burattin2014}.

\subsection{Continuous Data Generation}

An event stream differs from an event log in two fundamental aspects. First of all, and event stream has not a predefined end (i.e., the user can generate as many events he wants, so the simulation can last for an unspecified amount of time). The second distinction consists in keeping the events sorted by their time, and not grouped.

\begin{figure}
	\centering
	\includegraphics[width=.7\textwidth]{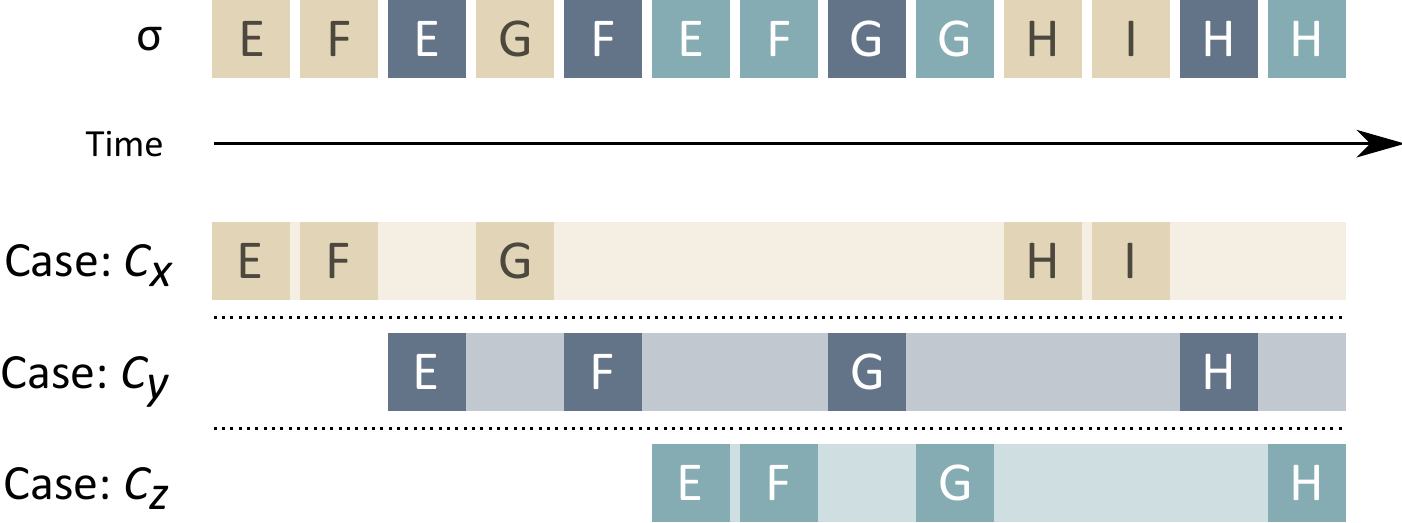}
	\caption{Graphical representation of an event stream. Boxes represent events: their background colors represent the case id, and the letters inside are the activity names. First line reports the stream, following lines are the single cases.}
	\label{fig:stream-example}
\end{figure}

Therefore, differently from an event log (which is a set of sequences, i.e., the traces), an event stream is just a sequence of events.
Therefore, the only property that must be enforced is that, given an event stream $\sigma$, for all indexes  $i$ available, $\#_\textit{time}(\sigma(i)) < \#_\textit{time}(\sigma(i+1))$. Instead, it will happen, for some indexes $i$, that $\#_\textit{case}(\sigma(i)) \neq \#_\textit{case}(\sigma(i+1))$ (i.e., contiguous events refer to different process instances). In this last case, the two events are said to belong to \emph{interleaving traces}.
Figure~\ref{fig:stream-example} reports a graphical representation of three interleaving traces and how the actual stream looks like.

From an implementation point of view, the idea is to create a socket, which is accepting connections from external clients. \plg, then, ``emits'' (i.e., writes on the socket) events that are generated. The challenge, in this case, is to let the system simulate and send events for a potentially infinite amount of time.

In order to generate our continuous stream, we need to ask the user for two parameters: the maximum number of parallel instances running at the same time, and the ``time scale''. The first parameter is used to populate the data structures required for the generation of the stream. Then, since the event emission is performed in ``the real time'' (opposed to the ``simulated time''), it might be necessary to scale the simulation time in order to have the desired events emission rate. To this end we need a time multiplier, which is expected to be defined in $(0, \infty]$. This time multiplier is used to transform the duration of a trace (and the time position of all the contained events), from the simulation time to the real time.

The procedure for the generation of streams is reported in Algorithm~\ref{alg:Stream}. It starts by allocating as many priority queues as the number of parallel instances of the stream (line~\ref{alg:Stream:queues}). These queues are basically used as events buffer. Then, the procedure starts a potentially infinite loop for the events streaming. At the beginning of this loop, the algorithm first needs to check whether the buffer contains enough events. If this is not the case (line~\ref{alg:Stream:buffer}), then a new process instance is simulated (line~\ref{alg:Stream:simulate}, using Algorithm~\ref{alg:Simulate}) and, after applying the time scale (line~\ref{alg:Stream:scale}), all its events are added to the event buffer (line~\ref{alg:Stream:enqueue}). Events are enqueue considering their time order (i.e., events with lower timestamps have higher priority).\footnote{Implementation details are skipped here, but some time manipulations are required in order to insert the new trace after all events already enqueued and keeping a certain amount of time from the last event.} Once the algorithm is sure about the availability of events, it extracts (and removes), from the buffer, the event with highest priority (line~\ref{alg:Stream:prio}). At this point, it is necessary to make happen the mapping between the simulation and the real time: the algorithm has to wait for a certain amount of time, in order to ensure the correct event distribution in the real time (line~\ref{alg:Stream:wait}). After such wait, the event can finally be emitted (line~\ref{alg:Stream:emit}), and all connected clients are notified.

Please note that, every time the algorithm has to repopulate the buffer, it asks the framework for the process which has to be simulated (line~\ref{alg:Stream:process}). This is a fundamental point: the user can change the process for the simulation, without stopping the current stream emission, and if such change occurs, a \emph{concept drift} will be observed. Concept drifts \cite{Maggi2013,Gama2010,Manning2008,Grossi2011} represent another important characteristic, which fundamentally differentiate event streams from event logs and, therefore, identify a requirement.
\begin{algorithm}[t!]
	\caption{Stream \label{alg:Stream}}
	\LinesNumbered
	\DontPrintSemicolon
	\Input{%
		$p$: the number of parallel instances \\
		$m$: time multiplier
	}
	\Comment{Initialization of the data structures}
	queues $\gets \emptyset$ \Comment*{This is an event buffer}
	\For{$i =1$ \emph{\textbf{up to}} $p$} {
		queues $\gets$ queue $\cup$ $\{$a new priority queue$\}$ \; \label{alg:Stream:queues}
	}
	$l	\gets \bot$ \;
	\For{\hspace{-.3em}\emph{\textbf{ever}}} {
		\Comment{Populate the event buffer}
		\If(\Comment*[f]{Here $|$queues$|$ is the sum of sizes of all queues contained. Although the inequality could be $<1$, we prefer to use $2p$ since these operations could be performed in a different concurred thread}){$|$\emph{queues}$| < 2p$ \label{alg:Stream:buffer}} {
			$\textit{proc} \gets$ the process to simulate \; \label{alg:Stream:process}
			$t \gets$ simulate a new trace for $\textit{proc}$ \Comment*{Alg.~\ref{alg:Simulate}}  \label{alg:Stream:simulate}
			scale the trace duration (and events times) according to $m$ \;  \label{alg:Stream:scale}
			distribute the events of $t$ (sorted by their time) to the queue with the highest priority of the last event \;  \label{alg:Stream:enqueue}
		}
		\Comment{The actual streaming}
		$e \gets$ extract (and remove) the event with highest priority from all queues \Comment*{From queues} \label{alg:Stream:prio}
		\If{$l \neq\bot$} {
			$w \gets \#_\textit{time}(e) - \#_\textit{time}(l)$ \;
			wait for $w$ time units \; \label{alg:Stream:wait}
		}
		$l \gets e$\;
		$\#_\textit{time}(e) \gets $ now \;
		emit $e$ \Comment*{To all connected clients} \label{alg:Stream:emit}
	}
\end{algorithm}

Please note also that, in order to have a more accurate mapping between simulation and real time, the implementation of the buffer population procedure (lines~\ref{alg:Stream:buffer}-\ref{alg:Stream:enqueue} of Alg.~\ref{alg:Stream}) can be executed in an external thread.\footnote{This cannot ensure a \emph{completely correct} mapping, however the difference has empirically seen negligible.}

\begin{figure}[t]
	\centering
	\subfloat[Process model used for performance computation of the stream reported in this section. \label{fig:model-stream}]{\includegraphics[width=\textwidth]{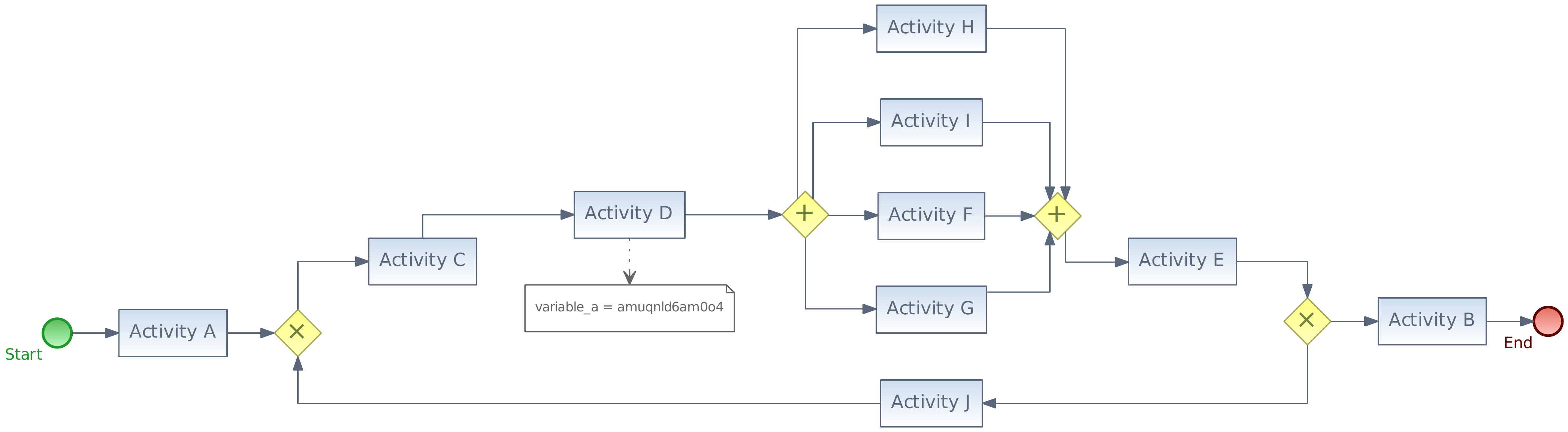}} \\
	\subfloat[Size of the buffer and total number of events sent for the process reported in Fig.~\ref{fig:model-stream}. \label{fig:chart-buffer}]{\includegraphics[width=.8\textwidth]{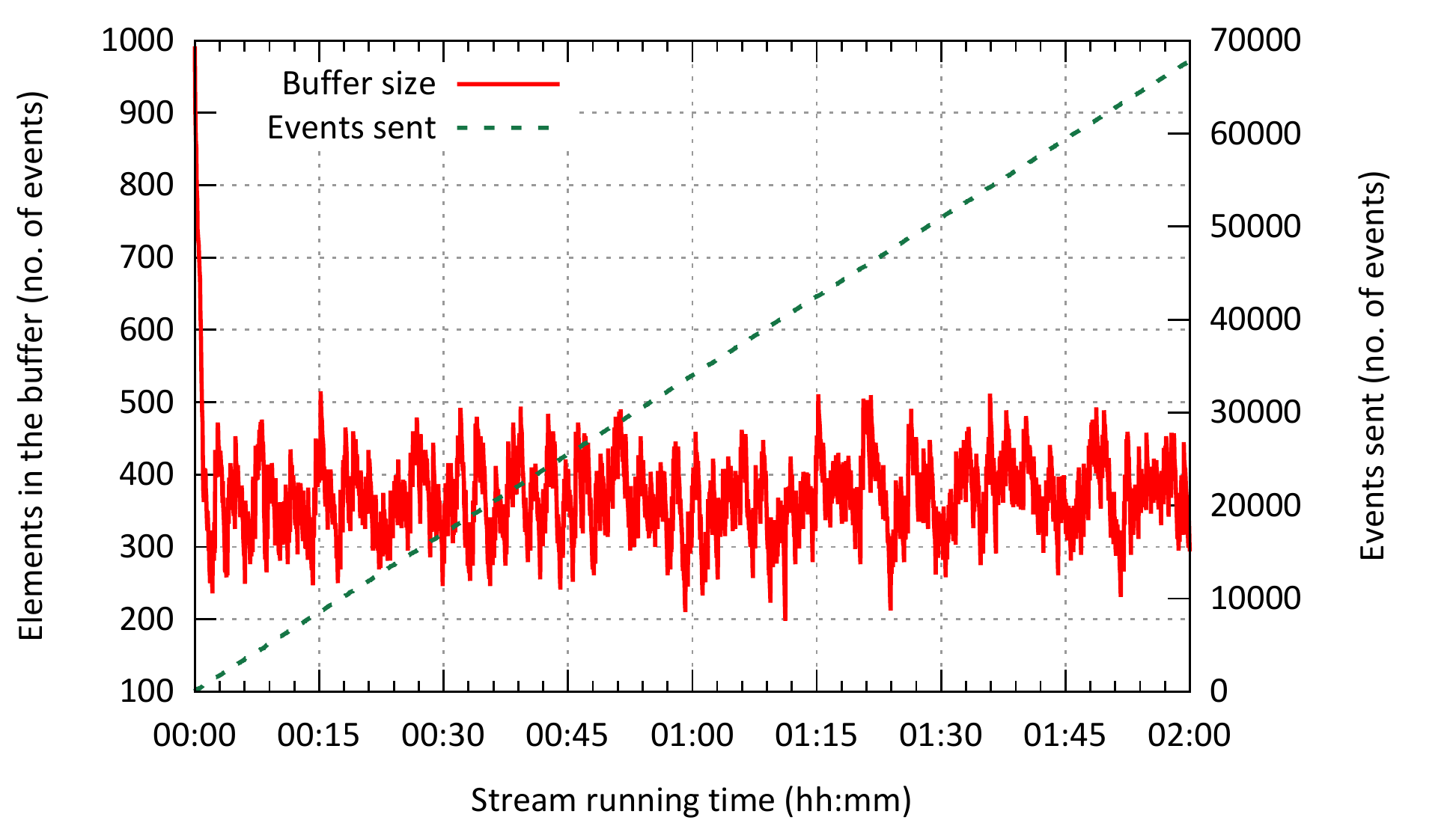}}
	\caption{Simulations details: the process model used and the size of the buffer. The entire simulation lasted for two hours.}
\end{figure}

\vspace{1em}
In order to assess the feasibility of this algorithm we run several experiments. In particular, we generated the process model reported in Fig.~\ref{fig:model-stream}. This process contains 10 activities, one parallel execution, one loop and one generated data object.
%
%\begin{figure}
%	\includegraphics[width=.5\textwidth]{img/models/model-test-stream.pdf}
%	\caption{Process model used for performance computation of the stream reported in this section.}
%	\label{fig:model-stream}
%\end{figure}
%
Then, we run the streamer of this process for two hours, generating, in total, $4\,174$ different traces and $67\,856$ events. The average throughput of the streamer, after an initial configuration stage, was set at 9.4 events per second.
%
%\begin{figure}
%	\includegraphics[width=.5\textwidth]{img/plot.pdf}
%	\caption{Size of the buffer and total number of events sent for the process reported in Fig.~\ref{fig:model-stream}. The simulation lasted for two hours.}
%	\label{fig:chart-buffer}
%\end{figure}
%
Figure~\ref{fig:chart-buffer} reports the memory requirement of the approach: the evolution of the buffer size and the total number of events sent are plotted against the running time of the actual stream. As the plot shows, the average number of stored event is between 300 and 400 events, which represents an affordable memory requirement for any hardware configuration available nowadays.

\subsection{Concept Drifts for Process Models}

One common characteristic of online settings is the presence of concept drifts. As described in the previous section, the tool is able to dynamically switch the source generating the events. However, in order to change the stream source, a new model is required. To create another model, two options are available: one is to load or generate from scratch a model; the other is to ``evolve'' an existing one: this is an important feature of \plg.

To evolve an existing model, \plg replaces an activity with a subprocess generated using the context-free grammar described in Section~\ref{sec:grammar}. This operation, which takes place randomly, and with a probability provided by the user, is repeated for each activity of the process. The new process could be very similar to the originating one, or very different, and this basically depends on the probability configured by the user.

\begin{figure}[t!]
	\centering
	\subfloat[Starting process model, randomly generated.]{\includegraphics[width=.9\textwidth]{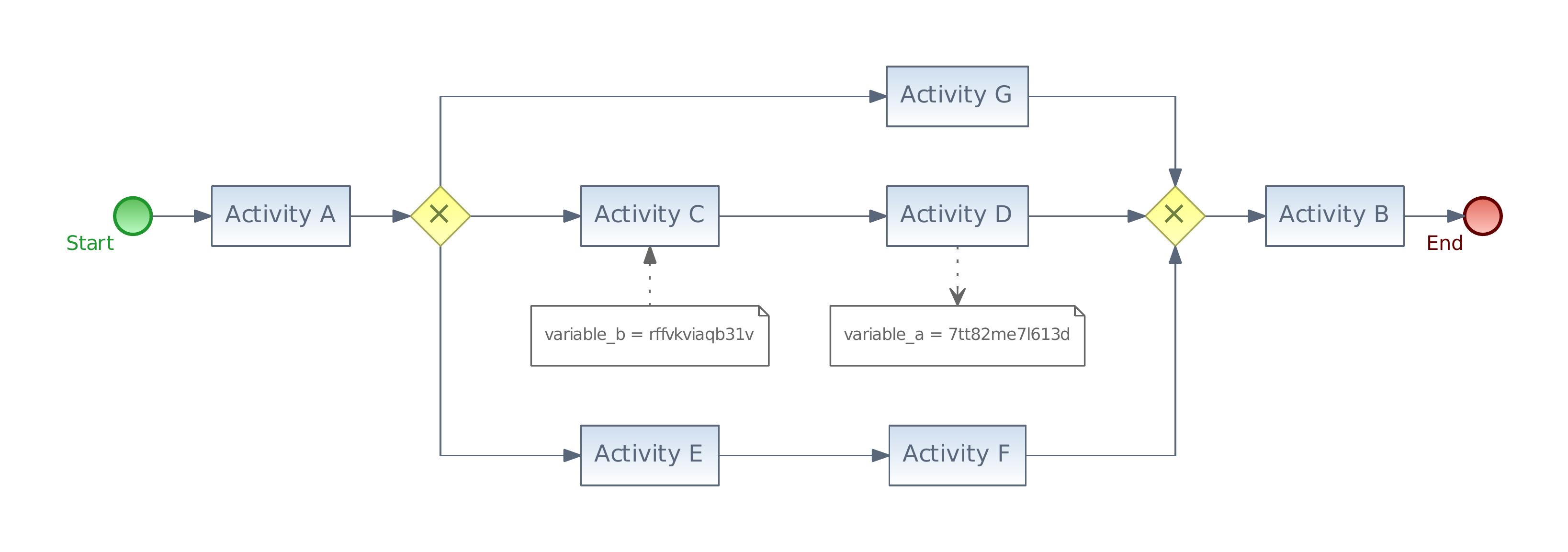}}\\
	\subfloat[First evolution of the process model. In this case, activity G has been replaced by a sequence of three activities (H, I, J).]{\includegraphics[width=.9\textwidth]{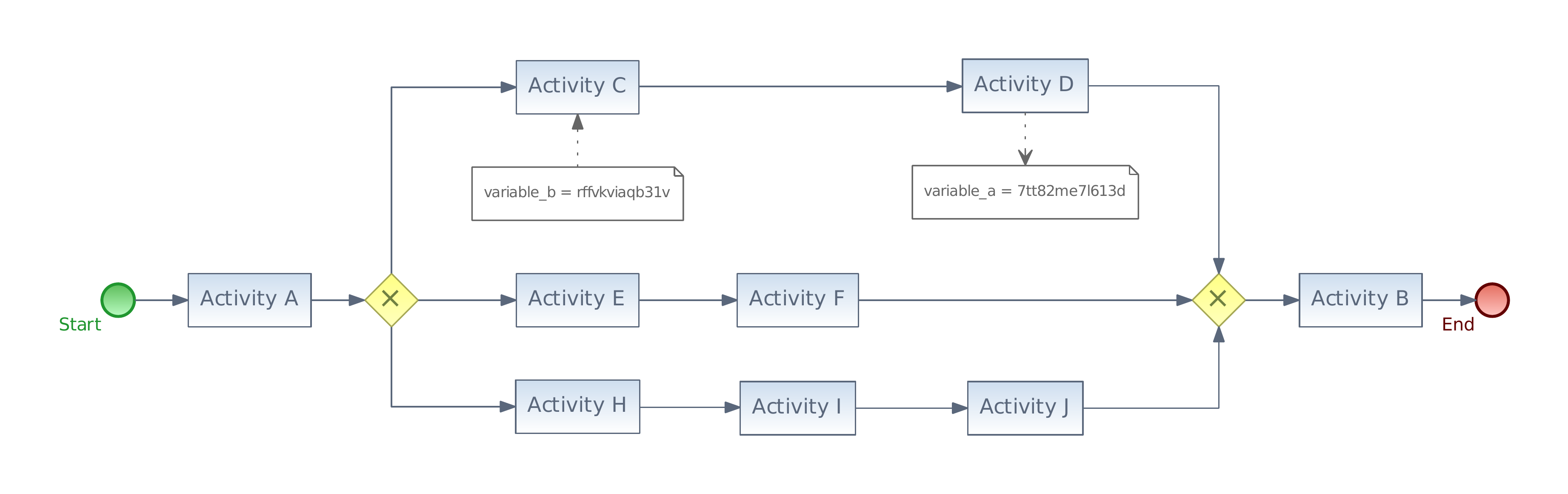}}\\
	\subfloat[Second evolution of the process model. In this case activity D --- and the associated data object `\texttt{variable\_a}' --- have been removed.]{\includegraphics[width=.9\textwidth]{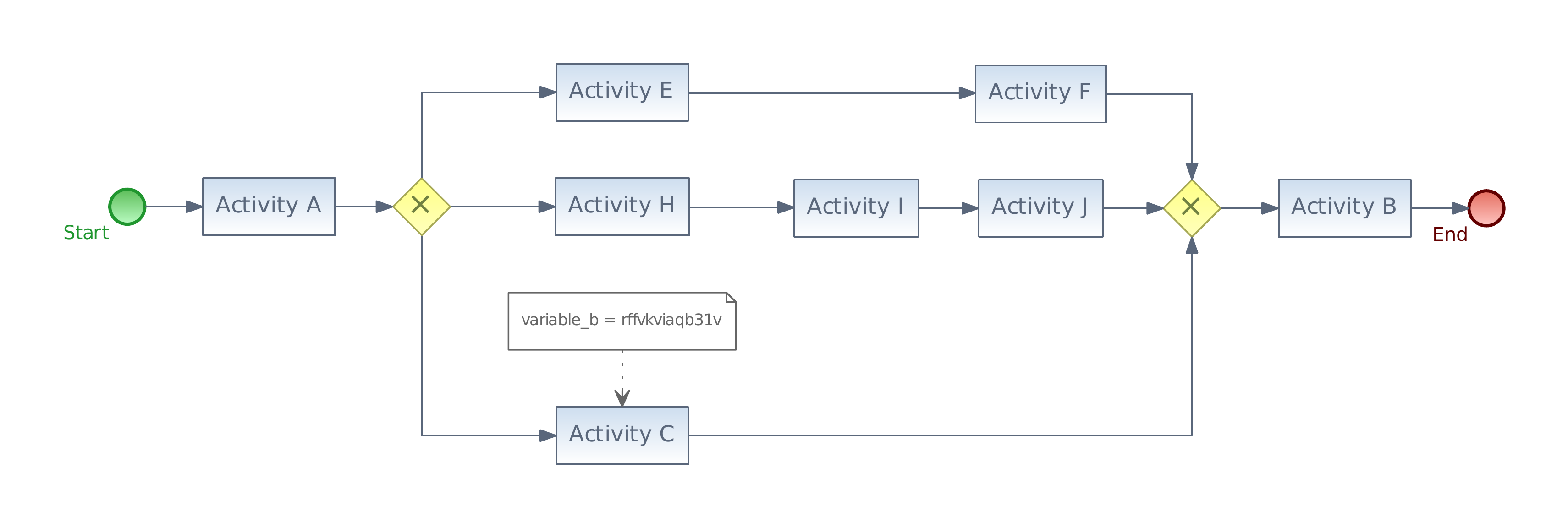}}
	\caption{A process model randomly generated with two sequential evolutions. Please note that new activities can be introduced or removed (with the associated data objects).}
	\label{fig:evolutions}
\end{figure}
For example, Figure~\ref{fig:evolutions} reports two evolutions of the process model, which has been randomly generated and which is reported in (a). In (b) the procedure applies the evolution by replacing ``Activity G'' with the sequence of three activities (``Activity H'', ``Activity I'' and ``Activity J''). In (c), the evolution involves ``Activity D'' (and the associated data object) which is replaced with a skip (i.e., it is removed).

Please note that an evolution could involve the creation or the deletion of data objects as well. Process evolution, therefore, can be used for the definition of particular experiments (e.g., a stream with random concept drifts occurring every 1000 events).

\section{Implementation Details of \plg}
\label{sec:implementation}

\plg has been implemented in a Java application. It is available as open source project and also binary files are provided for convenience.\footnote{See \url{http://plg.processmining.it} and \url{https://github.com/delas/plg}.} The project APIs can also be easily used to randomly generate processes or logs.
\begin{lstlisting}[float,style=customJava, caption={Java fragment for the creation of a new process, its simulation (to generate 1000 traces), and its export as Petri net.}, label=lst:java-api]
// process randomization
Process p = new Process("test");
ProcessGenerator.randomizeProcess(p,
                         RandomizationConfiguration.BASIC_VALUES);

// log simulation to generate 1000 traces
XLog l = new LogGenerator(p,
                 new SimulationConfiguration(1000)).generateLog();

// export as pnml
new PNMLExporter().exportModel(p, "p.pnml");
\end{lstlisting}
Listing~\ref{lst:java-api} reports the Java code required to generate a random process model; to simulate it in order to create 1000 traces; and to export the process as a Petri net (using the PNML standard).

The current implementation is able to load BPMN files generated with Signavio or \plg. It is also possible to export a model as PNML \cite{Hillah2009} or \plg file. Moreover, it is possible to export graphical representation of the model (both in terms of BPMN and Petri net) using the Graphviz file format \cite{Ellson2004}. The simulation of log files generates a XES\footnote{See \url{http://www.xes-standard.org}.}-compliant \cite{Gunther2009} objects, which can be exported both as (compressed) XES or (compressed) MXML. These formats are widely used by most process mining tools.

\begin{figure*}
	\includegraphics[width=\textwidth]{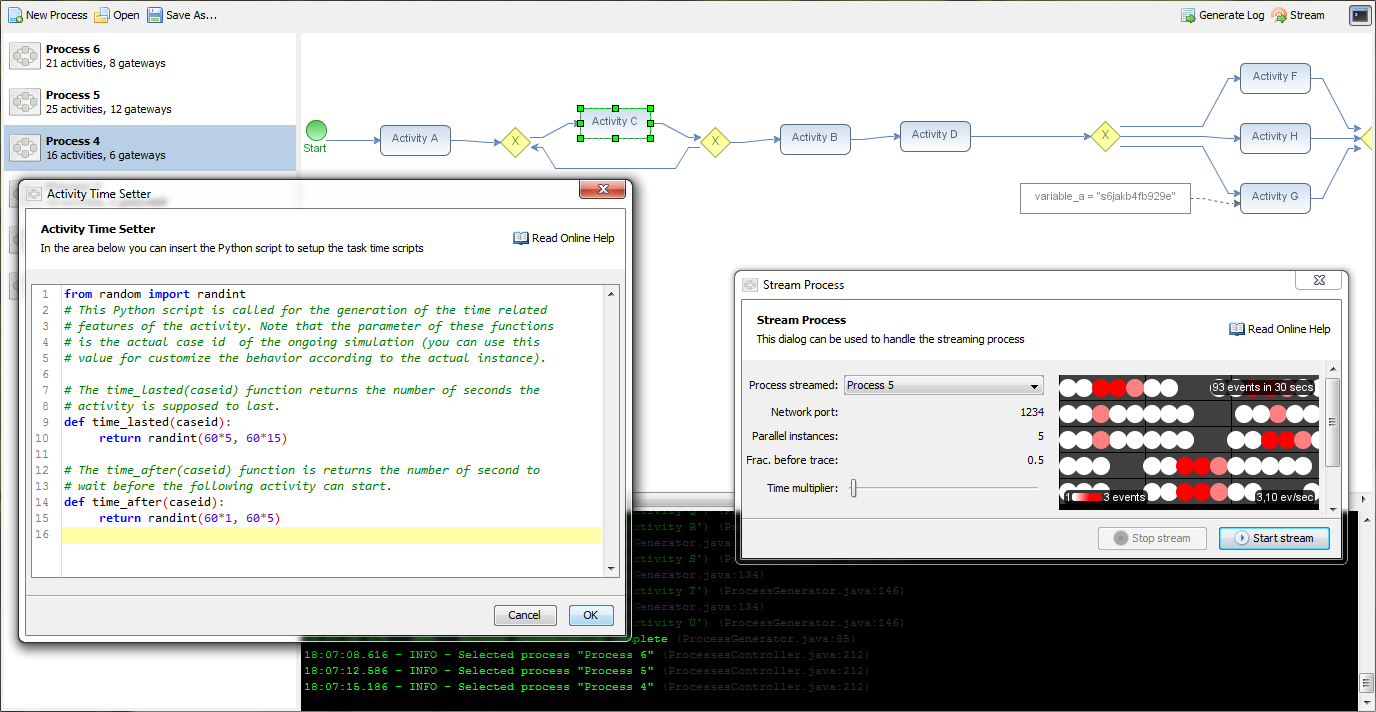}
	\caption{Screenshot of \plg. From the current visualization it is possible to see that several models have been created in the workspace, the dialog for the time rules configuration for ``Activity C'', and the console, which shows general information on what is going on. The stream dialog is reported as well.}
	\label{fig:screenshot}
\end{figure*}

Figure~\ref{fig:screenshot} reports a screenshot of the current implementation of \plg.
From the picture it is possible to see the main structure of the GUI: there is a ``workspace'' list of generated processes on the left. The selected process is shown on the main area. Right clicking on activities allows the user to set up  activity-specific properties (such as times, or data objects). On the bottom part of the main application it is possible to see the \plg console. Here the application reports all the log information, useful for debugging purposes. The application dialog in the foreground is used for the configuration of the Python script which will be used to determine the time properties. As shown, specific syntax highlighting and other typing hints (such as automatic indentation) helps the user in writing Python code. The stream dialog is also displayed in foreground. As can be seen, in this case, it is possible to dynamically change the streamed process and the time multiplier. The right hand side of such dialog (in the rectangle with black background), moreover, reports ``a preview'' of the stream: 30 seconds of the stream are reported (each round dot represents, in this case, up to 3 events).

As stated previously, some components of \plg require the execution of Python scripts. To deal with that we used the Jython framework\footnote{See \url{http://www.jython.org}.} which, basically, is an implementation of Python which can run in Java. The interaction between Java and Python objects is encapsulated in the \texttt{ScriptExecutor} hierarchy, reported in Fig.~\ref{fig:uml-class}.

\vspace{1em}
Since it is possible to repeat the code fragment reported in Listing~\ref{lst:java-api} as many times as required, we are able to fulfill \nameref{c1} (Section~\ref{sec:challenges}).
The detailed process simulation, the advanced data values generation and the noise configuration are necessary to create realistic multiperspective event logs and therefore to accomplish \nameref{c2}.
Finally, the feasibility of the stream procedure reported, together with its main features (such as the possibility to generate multiperspective streams, the dynamic change of the originating process model and the possibility to adapt the time between events emitted) makes possible to successfully cope with \nameref{c3}.

\section{Case Studies}
\label{sec:case-study}

In this section we would like to propose two possible scenarios in which \plg could easily be applied. In particular we will show a multiperspective analysis, performed in offline setting; and a control-flow discovery activity in online scenario.

\subsection{Offline Setting}

On the first case study, we would like to analyze both the control-flow and the data perspectives of a log files.

\begin{figure}
	\includegraphics[width=\textwidth]{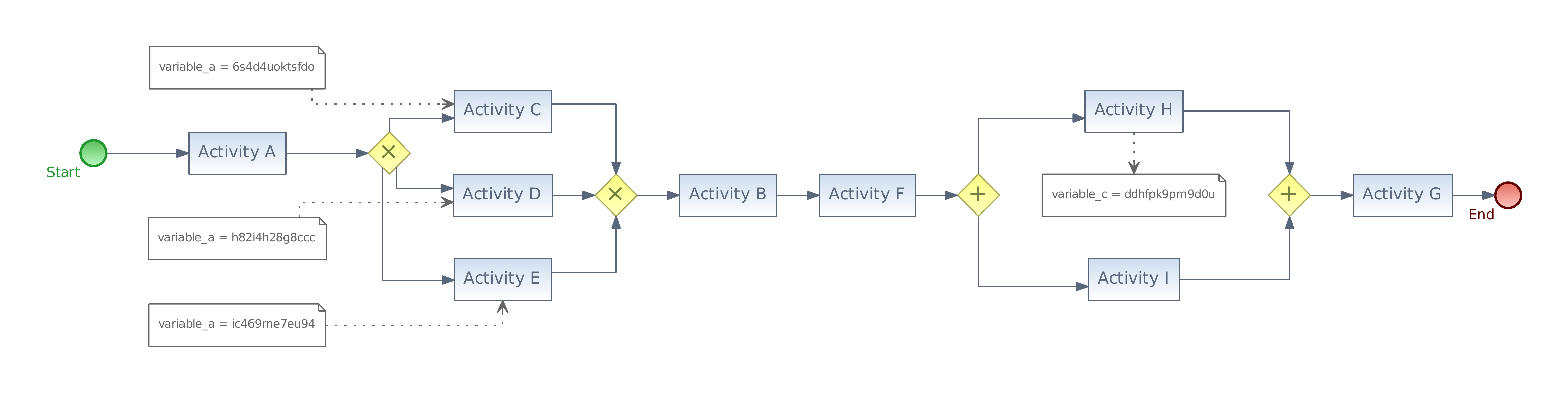}
	\caption{Process model used for the offline simulation.}
	\label{fig:offline-gs}
\end{figure}

To perform our test, we generated a random process model, and then we manually slightly modified it, in order to fit our goals. Specifically, we added required data objects to activities C, D, and E. These data objects are all named \texttt{variable\_a}, but each of them has a different value.
The generated model is reported in Figure~\ref{fig:offline-gs} and represents our gold standard.
Therefore, when we perform the data analysis, we expect the presence of a variable influencing the control-flow for those activities.

\begin{figure}[t]
	\centering
	\subfloat[Control-flow extracted using the Inductive Miner algorithm and then converted into a Petri net. \label{fig:offline-pn}]{\includegraphics[width=.9\textwidth]{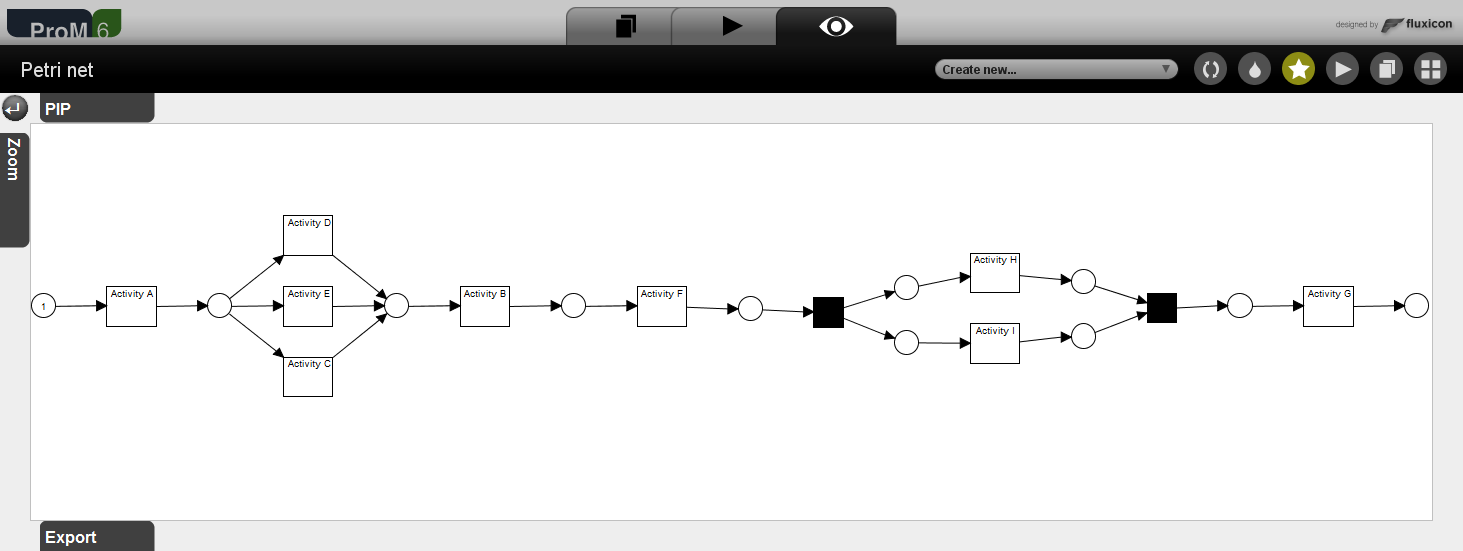}}\\
	\subfloat[Result of mining the data flow, which decorates the mined Petri net. \label{fig:offline-df}]{\includegraphics[width=.5\textwidth]{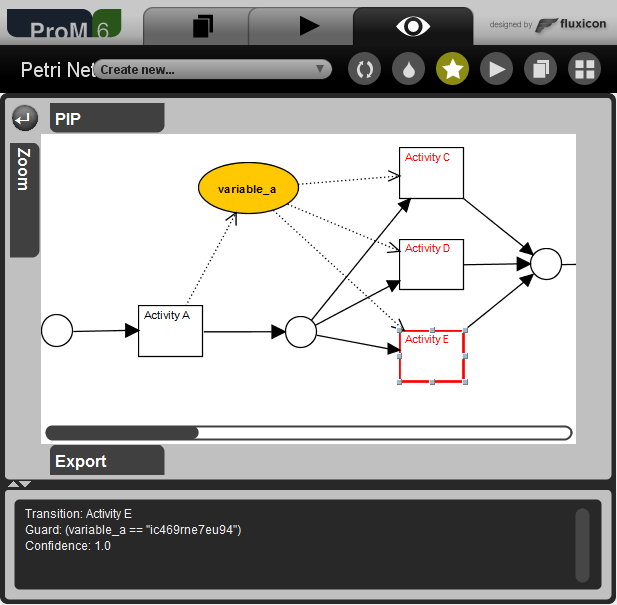}}
	\caption{Results of mining activities (control-flow and data flow) performed using ProM plugins.}
	\label{fig:offline}
\end{figure}

To perform our simulation, we generated a log with 2000 traces and then we analyzed it using ProM\footnote{See \url{http://www.promtools.org}.}~\cite{Verbeek2010}. For the control-flow discovery analysis we run the Inductive Miner~\cite{Leemans2013} algorithm. Then, we converted the generated model into a Petri net. The result is reported in Figure~\ref{fig:offline-pn}. As we can see, from the behavioral point of view, the mined model reflects the original one, except for the data perspective (which cannot be extracted with Inductive Miner).
Starting from the Petri net mined, we run the Data-flow Discovery plugin~\cite{DeLeoni2013a} in order to add data variables governing the control-flow. The result, which is reported in Figure~\ref{fig:offline-df}, shows the presence of a variable named \texttt{variable\_a} which is written by activity A, and read by activities C, D, and E. The screenshot also reports the actual guard for activity E (i.e., the value that is required in order to execute that activity).
Both the control-flow and data flow mined reflect the expected ones.

\begin{figure}
	\centering
	\includegraphics[width=.6\textwidth]{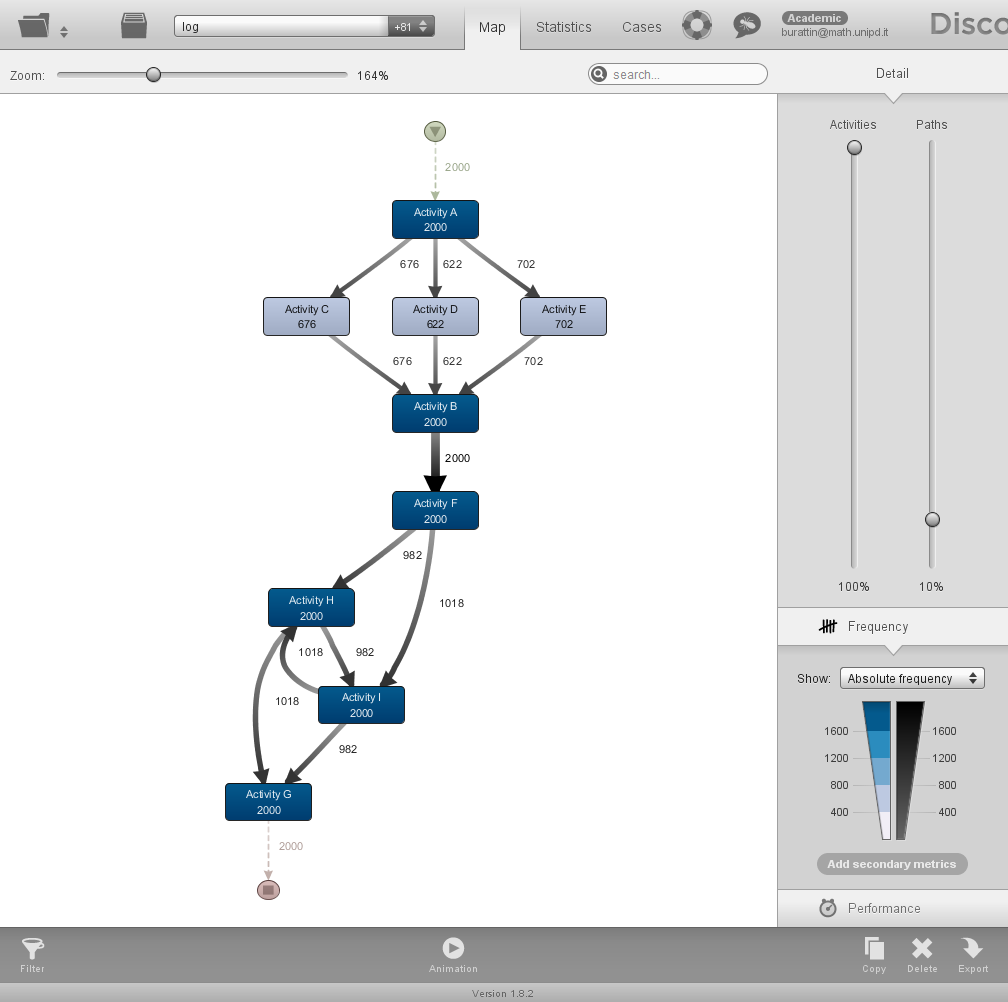}
	\caption{Result of mining the log using the tool Disco.}
	\label{fig:disco}
\end{figure}

As a second test, we mined the control-flow using the tool Disco.\footnote{See \url{http://fluxicon.com/disco/}.} The control-flow discovered by the tool is shown in Figure~\ref{fig:disco}.
The formalism, adopted by the tool for the representation of business processes, allows us to see, basically, only direct following relationships. As we can note, activities C, D and E are executed, respectively 676, 622, and 702 times. Since in total we have 2000 traces, this is an indication that maybe those activities are mutually exclusive (although this is not necessary). Instead, activities H and I are both executed 2000 times but we see there are connections between them. These connections indicate that the activities are not executed in a specific order (i.e., they are parallel). These behavioral characteristics reflect the gold standard.

\subsection{Online Setting}

For the second case study, we decided to analyze the online scenario with concept drifts. To achieve this goal, we created a second model ($M_2$), different from the previous one ($M_1$). Then we started streaming events referring to $M_1$.

\begin{figure}[t]
	\centering
	\subfloat[First model (after 500 events).]{\includegraphics[width=.48\textwidth]{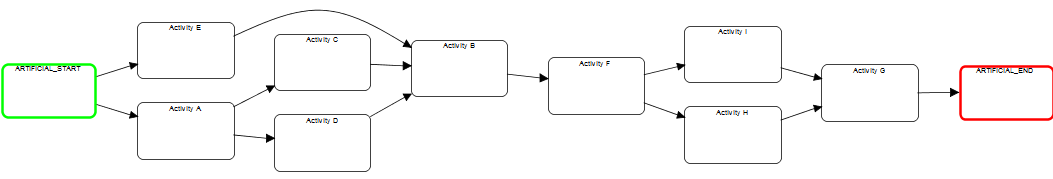}} \hfill
	\subfloat[Intermediate model.]{\includegraphics[width=.48\textwidth]{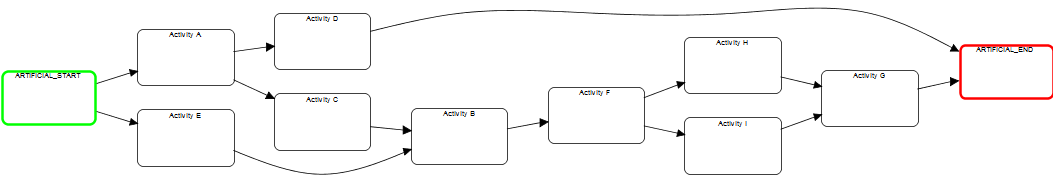}} \\
	\subfloat[Correct first model (after 2000 events).\label{fig:stream-miner:goal1}]{\includegraphics[width=.45\textwidth]{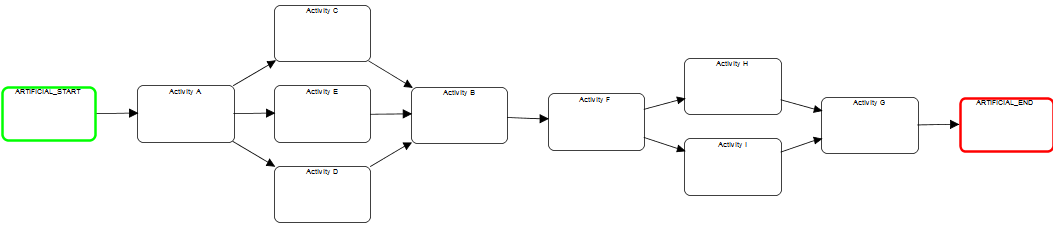}} \hfill
	\subfloat[Model mined after the drift occurred.\label{fig:stream-miner:afterdrift}]{\includegraphics[width=.45\textwidth]{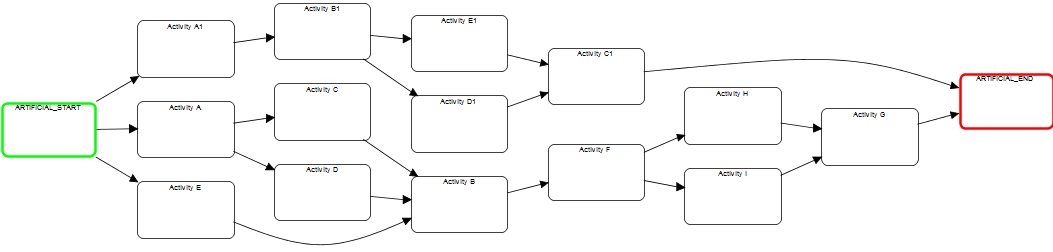}} \\
	\subfloat[Intermediate model.]{\includegraphics[width=.48\textwidth]{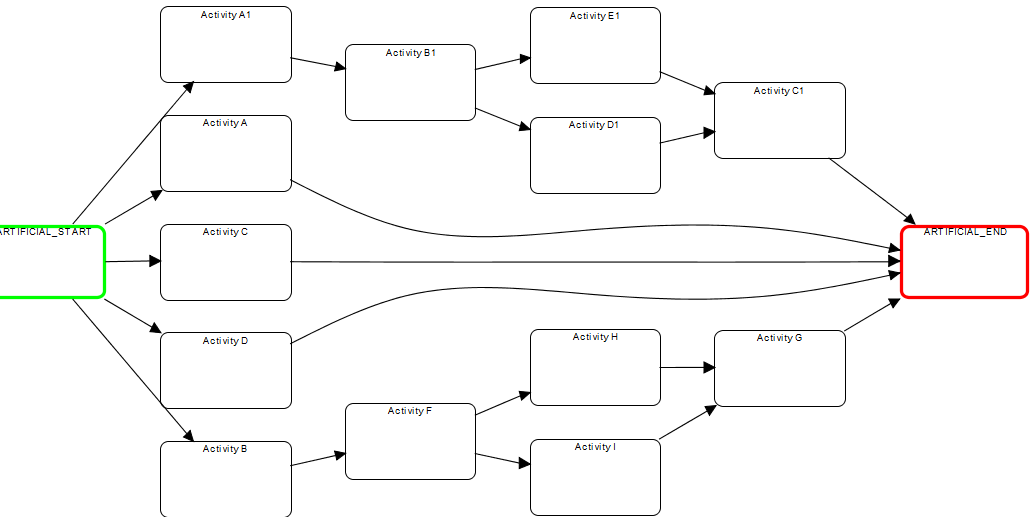}} \hfill
	\subfloat[Intermediate model.]{\includegraphics[width=.48\textwidth]{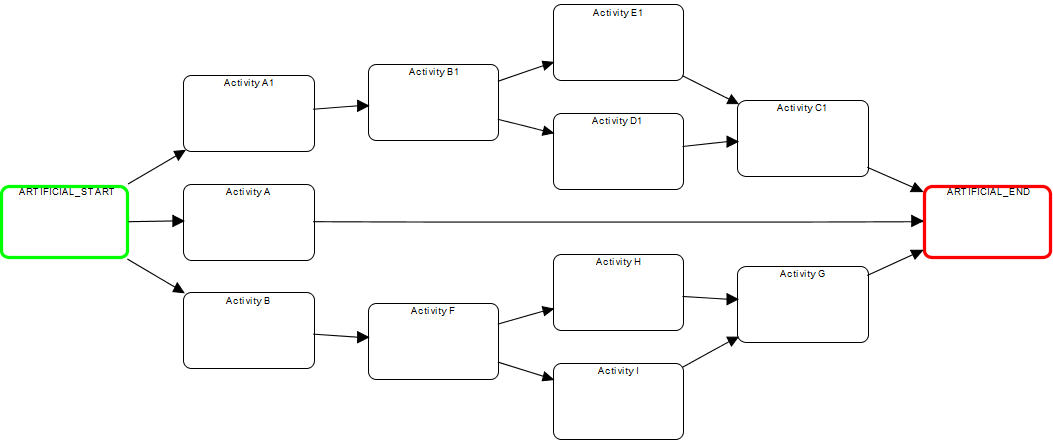}} \\
	\subfloat[Correct second model extracted (about 10000 events after the drift occurred).\label{fig:stream-miner:goal2}]{\includegraphics[width=.48\textwidth]{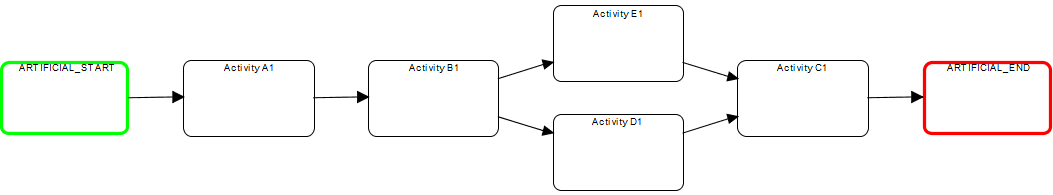}}
	\caption{Evolution of discovered models during a process stream simulation with a concept drift occurred.}
	\label{fig:stream-miner}
\end{figure}

In the meanwhile, we configured the stream mining plugin implemented in ProM and described in~\cite{Burattin2014}. Specifically, we used the mining approach based on Lossy Counting, with parameter $\epsilon = 0.032$. We also configured the miner to update the graphical representation of the process model every 500 events received. The sequence of models extracted is reported in Figure~\ref{fig:stream-miner}.

The first two models extracted are not equivalent to the expected one, since the miner needs several observations in order to reinforce and accept the patterns. Figure~\ref{fig:stream-miner:goal1} shows the model which is equivalent to the gold standard (in this picture split/join semantics are not reported for readability purposes). At this point, we decided to change the stream, and emit events referring to $M_2$ (i.e., we simulated the occurrence of a concept drift).
The first model extracted after such concept drift is reported in Figure~\ref{fig:stream-miner:afterdrift} and shows both process models $M_1$ and $M_2$ embedded into the same representation. This is a known phenomenon, and is due to the \emph{inertia} of the stream-based approaches.
After some events, since the miner is not receiving anymore observations from $M_1$, it starts to forget its structures. After some more events, the second model $M_2$ is definitely discovered, as shown in Figure~\ref{fig:stream-miner:goal2}, and no traces of $M_1$ are left.

\vspace{1em}
With these two case studies, we tried to show some of the possible usages of the described approaches. In these tests, we just used algorithms already available in the literature. However, the primary goal should be testing new ones.
Moreover, in the described cases, we just manually compared the mined models and the expected ones but this could be done automatically.
Finally, since we provide libraries to perform all functionalities via Java code, batch approaches could be designed, in order to perform the same operations against large repositories with models expressing very different behaviors.

\section{Conclusions and Future Work}
\label{sec:conclusion}

This paper describes \plg, which is the evolution of an already available tool. The old tool was able to randomly generate process models and simulate them. The new tool introduces updates on two sides: on one hand it extends the support to multiperspective models (by adding detailed control of time perspective and introducing data objects); on the other hand, full support for the simulation of online settings (generating drifting models and simulating event streams) is provided.

We believe, that the combination of the two newly introduced aspects allows the tool to be a valid instrument for the data mining, information systems, and process mining community, since it allows the simulation of very complex scenarios.
As the predecessor of this tool has proven, by its wide adoption, we think that the new features of \plg are important in order to push and help researchers to tackle the new challenges that upcoming settings propose (for example, \emph{big data} requires to handle streams of multiperspective data).

\vspace{1em}
We think that a lot of work is necessary in this field: the simulation of real scenarios is a very tough and broad task. In particular, it is important to investigate how to generate even more realistic scenarios. To achieve such realism, it is necessary to work both on the model generation (control-flow, time and data perspective) and on the simulation (for example identifying new types of noise). For example, the introduction of noise on the modeling could be considered (e.g., inserting or removing edges randomly, or in specific contexts).

An example of possible future work consists in the ad hoc simulation of the social perspective (identifying common patterns and possible behaviors) which, right now, is already possible, but just through the data perspective (e.g., generating data that describe the originators). Another future work, on the simulation part consists in introducing noise referring not to the trace/event modification, but to the distribution of the cases (i.e., not all control-flow paths are equally probable).

\bibliographystyle{abbrv}
\bibliography{library}

\end{document}